\documentclass[
  journal=largetwo,
  manuscript=research-article,
  year=2026,
  volume=00,
]{cup-journal}

\usepackage{aas-macros}
\usepackage{amsmath,amssymb}
\usepackage[nopatch]{microtype}
\usepackage{booktabs}
\usepackage{hyperref}
\usepackage[
    mode=match,
    range-units=single,
    range-phrase=\text{--},
    retain-explicit-plus
]{siunitx}
\usepackage[labelfont={sf,bf},font={sf,footnotesize}]{subcaption}
\usepackage{longtable}

\usepackage[australian]{babel}
\usepackage{csquotes}
\hypersetup{colorlinks, citecolor=blue, linkcolor=blue, urlcolor=blue}

\DeclareSIUnit{\yr}{yr} 
\DeclareSIUnit{\h}{h} 
\DeclareSIUnit{\pc}{pc} 
\DeclareSIUnit{\TB}{TB} 
\DeclareSIUnit{\mJy}{mJy} 
\DeclareSIUnit{\Jy}{Jy} 
\DeclareSIUnit{\kJy}{kJy} 

\title{Pulsar searches of \textit{Fermi}-LAT gamma-ray sources with the MWA}

\author{C.~P.~Lee} 
\affiliation{International Centre for Radio Astronomy Research, Curtin University, Kent Street, Bentley, WA 6102, Australia}
\email[C.~P.~Lee]{christopher.lee@icrar.org}

\author{N.~D.~R.~Bhat} 
\affiliation{International Centre for Radio Astronomy Research, Curtin University, Kent Street, Bentley, WA 6102, Australia}

\author{B.~W.~Meyers} 
\affiliation{Australian SKA Regional Centre (AusSRC), Curtin University, Kent Street, Bentley, WA 6102, Australia}
\alsoaffiliation{International Centre for Radio Astronomy Research, Curtin University, Kent Street, Bentley, WA 6102, Australia}

\author{W.~van~Straten} 
\affiliation{Manly Astrophysics, 15/41-42 East Esplanade, Manly, NSW 2095, Australia}

\author{D.~A.~Smith} 
\affiliation{Laboratoire d’Astrophysique de Bordeaux, Universit\'{e} de Bordeaux, CNRS, B18N, all\'{e}e Geoffroy Saint-Hilaire, F-33615 Pessac, France}


\begin{document}

\begin{abstract}

Searches of unassociated gamma-ray sources in the \textit{Fermi}-LAT catalogues have led to the discoveries of around a fifth of all known millisecond pulsars (MSPs).
These searches have almost exclusively been performed at radio frequencies above \qty{300}{\MHz}, where dispersion and scattering in the interstellar medium are less significant.
We report on a shallow survey for pulsars targeting 308 unassociated \textit{Fermi}-LAT sources in archival Murchison Widefield Array (MWA) observations from the Southern-sky MWA Rapid Two-metre (SMART) pulsar survey at \qty{154}{\MHz}.
This is the largest radio survey of unassociated \textit{Fermi}-LAT sources to date, and only the second to be conducted below \qty{300}{\MHz} after a survey with the Low Frequency Array (LOFAR) that discovered three MSPs.
Each source was observed for \qty{20}{\minute} by digitally beamforming the MWA tile voltages.
Searches were then performed using a new pipeline that implements a semi-coherent dispersion removal scheme for MWA data, enabling greater sensitivities to MSPs than is possible with fully-incoherent dispersion removal (e.g. 2--3 times better sensitivity for dispersion measures between \qtyrange{20}{40}{\per\cm\cubed\pc}).
The pipeline was tested by blindly detecting five known MSPs, four of which are in short-orbit binaries.
No new pulsars were identified in the survey, which we attribute to insufficient sensitivity.
We estimate flux density limits of approximately \qtyrange{30}{220}{\mJy} at \qty{154}{\MHz} (or \qtyrange{0.7}{5.2}{\mJy} at \qty{1.4}{\GHz}) for a spin period of \qty{2}{\ms} and a duty cycle of \qty{28}{\percent}, with a dependence on the sky temperature and the offset from the phase-centre of the primary beam.
We discuss how the improved instantaneous sensitivity from the Phase III upgrade of the MWA will increase the number of detectable gamma-ray pulsars by $\sim$\qty{30}{\percent} for the same integration time.
Additionally, the real-time beamformer (under development) will enable longer observations with sensitivities that are more competitive with previous surveys of \textit{Fermi}-LAT sources.
The semi-coherent search pipeline we have developed will also be useful for searches of supernova remnants, globular clusters, and pulsar candidates identified in imaging surveys, all of which will help to inform the significance of future surveys with SKA-Low.

\end{abstract}
\section{Introduction}

Millisecond pulsars (MSPs) are extremely valuable astrophysical tools due to the precision with which their pulses can be timed in both radio and gamma-rays \citep[e.g. arrival time estimates with $\sim$\unit{\us} uncertainties;][]{Spiewak2022,Valtolina2026}.
Around one in five of the over 600 Galactic MSPs\footnote{\url{https://www.astro.umd.edu/~eferrara/pulsars/GalacticMSPs.txt}} discovered to date have been found in targeted radio searches of unassociated gamma-ray sources detected by the Large Area Telescope (LAT) onboard \textit{Fermi} \citep{Ray2012}.
Almost all of these surveys have been conducted at radio frequencies above \qty{300}{\MHz}; either at \qty{1.4}{\GHz} \citep[e.g.][]{Cognard2011,Barr2013,Camilo2015,Clark2023}, \qty{820}{\MHz} \citep[e.g.][]{Ransom2011,Thongmeearkom2026}, or \qty{350}{\MHz} \citep[e.g.][]{Hessels2011,Cromartie2016,Bangale2024}.
These surveys have been particularly effective in discovering `spider' systems, which consist of a pulsar in a compact binary with a low-mass companion, often exhibiting radio eclipses.
Spider pulsars can be used to study binary pulsar evolution, particle acceleration, and measure neutron star masses \citep[see][for a summary]{Koljonen2025}.
Furthermore, 25 MSPs discovered in \textit{Fermi}-guided radio searches have since been included in pulsar timing arrays (PTAs), increasing the sensitivity to low-frequency gravitational waves \citep[e.g.][]{Siemens2013}.
The fourth \textit{Fermi}-LAT source catalogue \citep[4FGL;][]{4FGL} contains 1336 unassociated gamma-ray sources, many of which are likely to be pulsars.
A recent population synthesis by \citet{Sautron2025} predicted that there remain up to 220 unidentified pulsars to be discovered in the 4FGL catalogue.
There is also evidence suggesting that the fastest MSPs tend to be detected in gamma-rays and have unusually steep radio spectra \citep[e.g.][]{Espinoza2013,Bassa2017ApJ}.
There is therefore ample motivation to continue searching these sources, particularly at frequencies below \qty{300}{\MHz} and above \qty{2}{\GHz} where fewer surveys have been performed and there may be detectable pulsars with steeper or shallower intrinsic flux-density spectra.

Surveys below \qty{300}{\MHz} face several challenges due to chromatic effects on the radio signals imparted by the cold ionised interstellar medium (IISM).
Perhaps the most significant challenge is the computational cost of correcting the large dispersive delays over the receiver bandwidth.
Pulses propagating through the IISM experience a delay proportional to $\nu^{-2}$ (where $\nu$ is the observational frequency) and the dispersion measure (DM), which represents the column density of free electrons along the path of propagation \citep[see Section~4.1.1 of][]{LorimerKramer2012}.
As a result, pulses arrive at the receiver later at lower frequencies than their higher frequency counterparts, causing temporal smearing in the recorded time series.
Pulsar surveys must search over a range of DMs to identify dispersed signals.
Dispersion can be partially corrected by channelising the received signal and applying the appropriate time delays so that each pulse arrives simultaneously in all channels (i.e. `incoherent' dedispersion).
However, without correcting the dispersion within the frequency channels (intrachannel dispersion), there will always be residual dispersive smearing.
This is a particularly significant issue when searching for short-period pulsars at low frequencies, where intrachannel dispersive smearing can limit searches to very low DMs.
Assuming the intervening medium is a cold tenuous plasma (as is the case for the IISM), intrachannel dispersion can be completely removed by phase-coherently dedispersing the complex voltages for each channel \citep{Hankins1971,vanStraten2011}.
However, this requires several computationally expensive steps that make fully-coherent searches largely impractical.
\citet{Bassa2017A&C} proposed an alternative `semi-coherent' dedispersion strategy, which involves performing phase-coherent dedispersion to a smaller number of DMs (e.g. $\Delta\mathrm{DM}\sim \qty{1}{\per\cm\cubed\pc}$), then incoherently dedispersing to a larger number of DMs in between each coherent step (e.g. $\Delta\mathrm{DM}\sim \qty{e-3}{\per\cm\cubed\pc}$).
By carefully choosing the DM step sizes, one can effectively trade off between computational cost and dispersive smearing.
This novel approach was used to discover three MSPs in a targeted survey of unassociated \textit{Fermi}-LAT sources with the Low-frequency Array (LOFAR) at \qty{135}{\MHz} \citep{Pleunis2017,Bassa2017ApJ,Bassa2018}.

\begin{figure*}[t]
    \centering
    \includegraphics[width=0.85\linewidth]{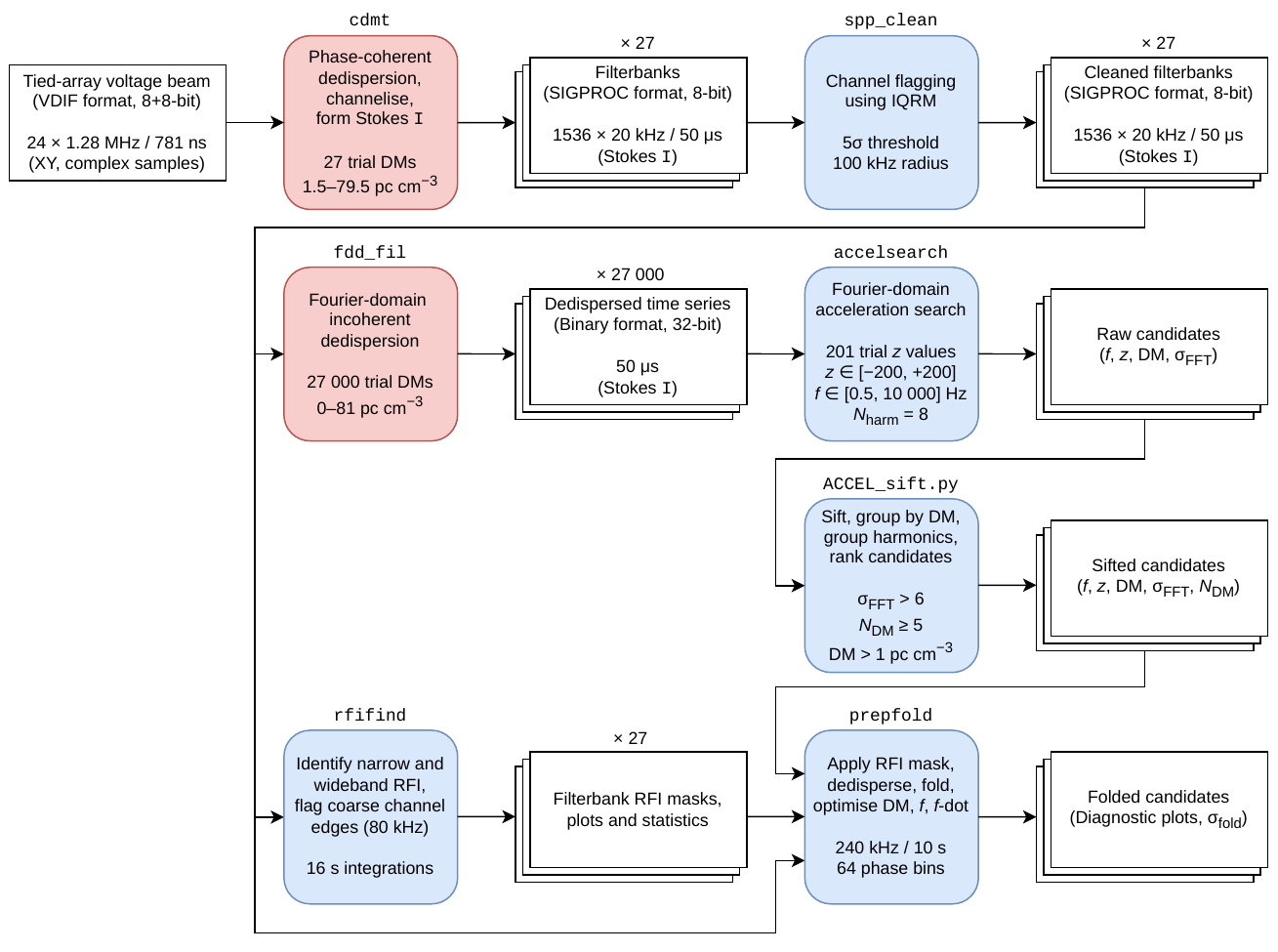}
    \caption{Diagram illustrating the basic workflow of the semi-coherent pulsar search pipeline. Red rounded boxes indicate GPU-based processing tasks and blue rounded boxes indicate CPU-based processing tasks. See Sections~\ref{sec:pipeline:description} and \ref{sec:ddplan} for details.}
    \label{fig:workflow}
\end{figure*}

The demonstrated benefits of semi-coherent dedispersion in the LOFAR survey motivated us to apply the technique to pulsar searches with the Murchison Widefield Array (MWA).
The MWA is a low-frequency aperture array telescope located at Inyarrimanha Ilgari Bundara, the CSIRO Murchison Radio-astronomy Observatory in Western Australia \citep{Tingay2013}.
It operates between \qtyrange{70}{300}{\MHz} and can observe declinations south of around \qty{+30}{\degree}.
It is currently the most sensitive telescope operating below \qty{300}{\MHz} with access to the entire southern sky.
The MWA's Voltage Capture System \citep[VCS;][]{Tremblay2015,Morrison2023} provides the capability to record the complex voltages from each of the tiles in the array \citep[$\sim$\num{128} tiles in Phase~II;][]{Wayth2018} over a \qty{30.72}{\MHz} observing bandwidth.
To exploit the flexibility of VCS data, the Southern-sky MWA Rapid Two-metre (SMART) pulsar survey was conceived, comprising 71 drift-scan observations that together cover the entire sky within the declination range of the telescope with a dwell time of \qty{80}{\minute} \citep{Bhat2023a,Bhat2023b}.
The primary objective of SMART is to conduct an untargeted all-sky survey for pulsars.
However, the dispersive smearing limits detections to $\mathrm{DM}\lesssim\qty{30}{\per\cm\cubed\pc}$ for MSPs \citep{Bhat2023a}.
Given the high computational cost of processing and searching through the SMART data set, it is also infeasible to perform orbital acceleration searches for the full survey.
This means that the survey is not sensitive to pulsars in short-orbit binaries, including the majority of MSPs that have been discovered in \textit{Fermi}-guided radio searches.

To improve sensitivity to MSPs and exploit the unexplored period/DM/acceleration parameter space in the SMART survey, we have developed a semi-coherent dedispersion pipeline intended for targeted MSP searches in MWA data.
These searches will complement LOFAR by targeting \textit{Fermi}-LAT sources in the southern hemisphere below \qty{300}{\MHz} for the first time.
In Section~\ref{sec:pipeline}, we describe the pipeline workflow and implementation, and validate it by blindly detecting known MSPs.
In Section~\ref{sec:survey}, we describe a pilot survey of unassociated \textit{Fermi}-LAT sources in the SMART data, including the target selection, sensitivity, and search results.
In Section~\ref{sec:discussion}, we discuss the limitations of the survey and the prospects for discovering MSPs with the MWA Phase III.
Lastly, in Section~\ref{sec:conclusions}, we conclude and provide recommendations for future targeted searches with the MWA.

\section{Search Pipeline}\label{sec:pipeline}

\subsection{Description}\label{sec:pipeline:description}

The MWA tile voltages recorded by the VCS are coherently combined into a tied-array beam using the GPU-accelerated beamforming software \textsc{vcsbeam} \citep[][Bhat~et~al. submitted]{Ord2019}.
The beamformed data are stored as 8+8-bit complex samples in the VLBI Data Interchange Format (VDIF)\footnote{\url{https://vlbi.org/vlbi-standards/vdif/}} with one data file and one header file for each of the $24\times\qty[number-unit-product=\text{-}]{1.28}{\MHz}$ coarse channels (i.e. receiver channels), with a native time resolution of \qty{781.25}{\ns}.
The VDIF data, along with a dedispersion plan, are provided as inputs to the search pipeline.
The pipeline workflow is summarised in Figure~\ref{fig:workflow}.

We perform the phase-coherent dedispersion using \textsc{cdmt} -- a GPU-accelerated software designed to efficiently dedisperse to multiple DMs using a convolving filterbank \citep{Bassa2017A&C}.
\textsc{cdmt} also performs channelisation and detection into Stokes $I$.
As the code was originally designed to read LOFAR data in HDF5 format, we have modified it to read VDIF data\footnote{\url{https://github.com/cplee1/cdmt}}.
The detected data for each coherent DM trial are stored as 8-bit \textsc{sigproc} filterbanks \citep{Lorimer2011} containing $1536\times\qty[number-unit-product=\text{-}]{20}{\kHz}$ channels with a time resolution of \qty{50}{\us}.
Each filterbank is then incoherently dedispersed using the GPU-accelerated Fourier domain dedispersion (FDD) algorithm implemented by \citet{Bassa2022} in a fork of the \textsc{dedisp} library\footnote{\url{https://github.com/svlugt/dedisp}}.
The dedispersed time series for each DM trial are stored as 32-bit floats in a binary file with an accompanying header file for compatibility with \textsc{presto} \citep{Ransom2001}.

We search for the signals of binary pulsars in the dedispersed time series using the \textsc{accelsearch} routine from \textsc{presto}, which implements the Fourier-domain acceleration search (FDAS) to identify pulsar candidates.
The FDAS is described in detail in \citet{Ransom2002}, and a summary is provided in \citet{Andersen2018}.
For the FDAS, we assume that the observation length $T$ is a sufficiently small fraction of the binary orbital period $P_\mathrm{orb}$, such that the orbital acceleration of the pulsar can be approximated as constant.
Each harmonic of the fundamental spin frequency $f_0$ would then exhibit a constant frequency derivative $\dot{f}$, causing the signal to drift through $z=\dot{f}T^2$ bins in each Fourier power spectrum (where $z$ is the Fourier frequency derivative).
In this case, the acceleration may be expressed as
\begin{equation}
    \alpha = \frac{\dot{f}c}{hf_0}= \frac{zc}{hf_0T^2},
\end{equation}
where $c$ is the speed of light and $h$ is the harmonic number ($h=1$ is the fundamental).
The FDAS involves generating a set of template filters for a range of trial values of $\dot{f}$, and correlating the templates with the complex-valued Fourier-transformed time series.
The Fourier components are converted to powers and normalised by a block running median, which removes red noise and allows for accurate estimation of signal significances.
The Fourier power spectra for the $\dot{f}$ trials are then combined to form a 2D plane of powers in $f$-$\dot{f}$ space.
Candidate signals are identified via threshold searching in the $f$-$\dot{f}$ plane and incoherently summing harmonics to increase the sensitivity.
As discussed in \citet{Andersen2018}, most MSPs with stellar-mass companions are detectable with $|z|<200$ if $T\lesssim0.1P_\mathrm{orb}$.
We search $z$ values from \num{-200} to \num{200} in steps of $\Delta z=2$ (i.e. 201 acceleration trials), and spin frequencies from \qty{0.5}{\Hz} to \qty{10}{\kHz} with up to $N_\mathrm{harm}=8$ summed harmonics.
We are therefore sensitive to the 8th harmonic of signals with $f_0<\qty{1.25}{\kHz}$ (or $P>\qty{0.8}{\ms}$).
To compare candidates over multiple DM trials, \textsc{accelsearch} calculates the probability that the summed harmonic power of a candidate is due to noise (corrected for the number of independent $f$ and $\dot{f}$ trials searched), and expresses it in terms of its equivalent Gaussian significance, $\sigma_\mathrm{FFT}$.

The \textsc{accelsearch} candidates for all of the DM trials are then grouped and sifted using \textsc{python} utilities provided by \textsc{presto}.
First, candidates are rejected if (1) $\sigma_\mathrm{FFT}<6$; (2) the candidate is detected in only 1 harmonic that has a normalised Fourier power less than \num{100}; (3) the candidate is detected in multiple harmonics, but all harmonics have a normalised Fourier power less than \num{3}; or (4) the candidate is dominated by a single high-power and high-order harmonic.
The candidates are then grouped by period and DM, and are rejected if they appear in less than \num{5} DM trials or if the best DM is less than \qty{1}{\per\cm\cubed\pc}.
Only the candidate from the DM with the highest $\sigma_\mathrm{FFT}$ is kept.
Finally, harmonically-related candidates are grouped and only the harmonic with the highest $\sigma_\mathrm{FFT}$ is kept.
The candidate list is then sorted by $\sigma_\mathrm{FFT}$.
For our searches, $\sim$\num{100} candidates typically remain per pointing after sifting.
Although the candidates with $\sigma_\mathrm{FFT}$ between 6 and 10 are mostly due to noise, pulsars with narrow duty cycles will often experience a `boost' in significance when folded and optimised, due to the limited number of harmonics summed in the FDAS \citep{Sengar2023}.
Since our searches are targeted, we can afford to fold candidates down to a lower $\sigma_\mathrm{FFT}$ to take advantage of this effect.

We then use the \textsc{prepfold} routine from \textsc{presto} to dedisperse and fold the filterbank data at the DM, $f$, and $\dot{f}$ of each candidate with 10-s subintegrations, \qty[number-unit-product=-]{240}{\kHz} subbands, and 64 phase bins.
\textsc{prepfold} determines the optimal values for the DM, $f$, and $\dot{f}$ by performing a fine grid search around the candidate values and maximising the reduced $\chi^2$ of the profile.
A candidate plot is then generated displaying the folded data and other diagnostics to aid with interpretation.
Similar to \textsc{accelsearch}, \textsc{prepfold} also estimates the probability that the pulsations are due to noise, and expresses it in terms of its equivalent Gaussian significance, $\sigma_\mathrm{fold}$.

Although the SMART survey data are generally exceptionally clean, it is still necessary to perform basic mitigation of radio frequency interference (RFI).
To reduce spurious \textsc{accelsearch} candidates caused by strong and persistent narrowband RFI, outlier frequency channels in the \textsc{sigproc} filterbanks are flagged using Inter-Quartile Range Mitigation \citep[IQRM;][]{Morello2022} with a threshold of $4\sigma$ and a radius of \num{5} channels (\qty{100}{\kHz}).
We use the implementation of IQRM in the \textsc{spp\_clean} utility provided by \textsc{sigpyproc3}\footnote{\url{https://github.com/FRBs/sigpyproc3} (v1.2.0)}.
Additionally, we create an RFI mask for each filterbank using the \textsc{rfifind} routine from \textsc{presto}.
We use time blocks of \qty{16}{\s} and explicitly mask \num{4} channels (\qty{80}{\kHz}) at the top and bottom edges of each coarse channel (i.e. \qty{12.5}{\percent} of the bandwidth).
The \textsc{rfifind} masks are applied by \textsc{prepfold} before folding, which reduces both impulsive broadband RFI and weaker narrowband RFI in the folded data.
\textsc{prepfold} also performs clipping of outlier samples at zero-DM; however, because the filterbanks are phase-coherently dedispersed, signals at zero DM will be subject to intrachannel dispersive smearing, making impulsive RFI less significant.

The ranked list of candidates, candidate plots, and \textsc{rfifind} results (which includes summary plots and statistics) are saved for each search.
The candidate plots are inspected by eye for persistent broadband pulsations and peaks in the reduced $\chi^2$ as a function of DM, spin period ($P$), and spin period derivative ($\dot{P}$), which are all features of real pulsar signals.
Candidates are generally rejected if the signal is impulsive or narrowband, has a DM close to zero, or has a spin period that is a harmonic of a known terrestrial signal.
In some cases, the \textsc{rfifind} results are used as a diagnostic to assess the presence of RFI in the data.

\subsection{Dedispersion Plan}\label{sec:ddplan}

\begin{figure}[t!]
    \centering
    \includegraphics[width=\linewidth]{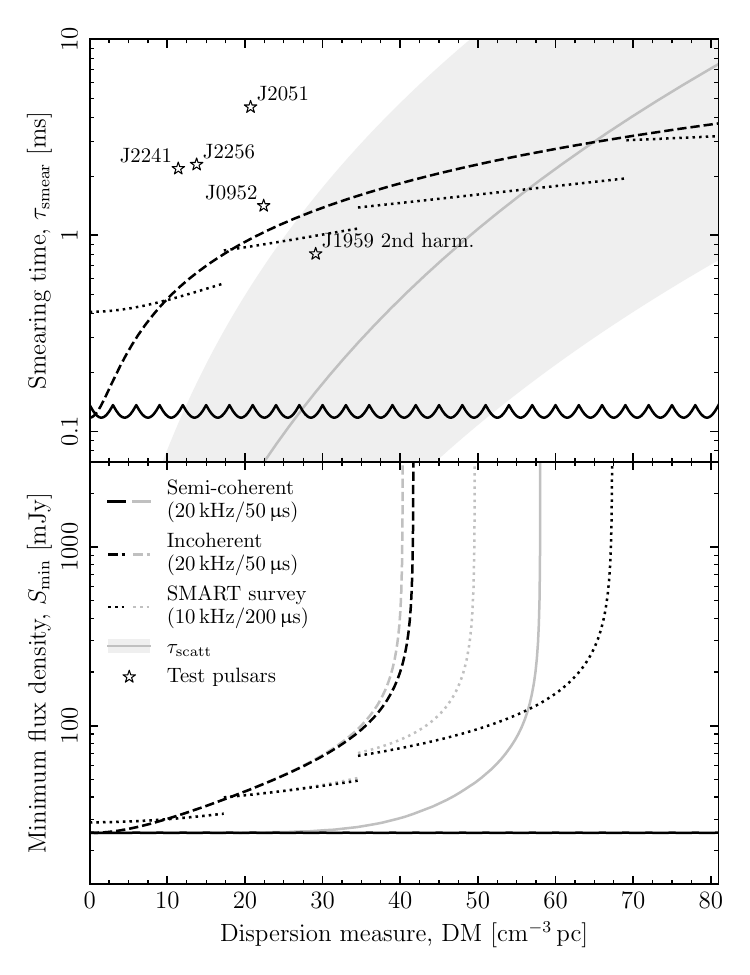}
    \caption{
    Comparison of dedispersion plans in the SMART frequency band (\qtyrange{138.88}{169.60}{\MHz}).
    \textit{Top:} The worst-case temporal smearing (i.e. $\tau_\mathrm{smear}$ assuming the maximum $\tau_{\delta\mathrm{DM}}$) as a function of DM.
    The estimated scattering time ($\tau_\mathrm{scatt}$) is shown with an order-of-magnitude error band \citep{Bhat2004}.
    The test pulsars are indicated at their respective spin periods and DMs, and labelled by their right ascension.
    \textit{Bottom:} The minimum detectable flux density ($S_\mathrm{min}$) for a pulsar with a spin period of \qty{2}{\ms} and a duty cycle of \qty{28}{\percent}, assuming an integration time of \qty{20}{\minute} and a SEFD of \qty{1}{\kJy} (i.e. MWA Phase II sensitivity away from the Galactic plane).
    The grey and black lines show $S_\mathrm{min}$ with and without scatter broadening, respectively.
    The solid line shows a semi-coherent dedispersion plan for a channel width of \qty{20}{\kHz}, a sample time of \qty{50}{\us}, an incoherent DM step size of \qty{0.003}{\per\cm\cubed\pc}, and a coherent DM step size of \qty{3}{\per\cm\cubed\pc}.
    The dashed line shows an equivalent dedispersion plan without coherent dedispersion.
    The dotted line shows the untargeted SMART survey dedispersion plan for a channel width of \qty{10}{\kHz} and a minimum sample time of \qty{200}{\us}; the discontinuities are due to the progressive downsampling of the sample time and DM step size to optimise the survey efficiency.
    }
    \label{fig:ddplan}
\end{figure}

The temporal smearing ($\tau_\mathrm{smear}$) of a search is the sum in quadrature of the sample time ($\delta t$), the intrachannel dispersive smearing\footnote{\label{fn:centrefreq}Calculated at the geometric centre frequency, $\sqrt{\nu_\mathrm{low}\nu_\mathrm{high}}$, where $\nu_\mathrm{low}$ and $\nu_\mathrm{high}$ are the lowest and highest frequencies of the observing band.} ($\tau_\mathrm{DM}$), and the dispersive smearing due to the difference between the DM of the pulsar and the nearest trial DM ($\tau_{\delta\mathrm{DM}}$):
\begin{equation}\label{eq:tsmear}
    \tau_\mathrm{smear} = \sqrt{\delta t^2 + \tau_\mathrm{DM}^2 + \tau_{\delta\mathrm{DM}}^2}.
\end{equation}
We have chosen a semi-coherent dedispersion plan that limits the temporal smearing to \qty{136.1}{\us} in the SMART frequency band for frequency/time resolutions of \qty{20}{\kHz}/\qty{50}{\us}.
The data are first phase-coherently dedispersed to \num{27} DMs from \num{1.5} to \qty{79.5}{\per\cm\cubed\pc} in steps of \qty{3}{\per\cm\cubed\pc}.
They are then incoherently dedispersed to $\pm\qty{1.5}{\per\cm\cubed\pc}$ in steps of \qty{0.003}{\per\cm\cubed\pc} around each coherent DM.
This results in a total of \num{27000} DM trials.
The worst-case smearing (i.e. when $\tau_{\delta\mathrm{DM}}$ is maximised) for the semi-coherent dedispersion plan is shown as a function of DM in Figure~\ref{fig:ddplan}, along with a fully-incoherent dedispersion plan and the untargeted SMART survey dedispersion plan for comparison.
We also show how the minimum detectable flux density scales with DM for a typical MSP, with and without pulse broadening due to scattering (see Section~\ref{sec:sensitivity} for details on the sensitivity calculations).

The maximum temporal smearing and DM range for our survey are similar to the LOFAR gamma-ray survey\footnote{The LOFAR survey had a maximum smearing of $\qty{149.5}{\us}$ and searched up to a DM of \qty{80}{\per\cm\cubed\pc}.} \citep{Pleunis2017}.
The Galactic DM contribution generally saturates within this range at high latitudes.
At lower latitudes and on the Galactic plane, where the DM may be higher, the sky temperature and interstellar scattering greatly limit sensitivity at higher DMs.
Indeed, we are yet to detect an MSP with a DM greater than \qty{80}{\per\cm\cubed\pc} with the MWA \citep{Lee2025}.

\subsection{Implementation and Benchmarking}
The pipeline is built using \textsc{nextflow}, which manages the flow of intermediate data products and the submission of jobs to the cluster \citep{DiTommaso2017}.
We have deployed the pipeline on the high-performance computing cluster at DUG Technology in Perth, which offers multiple node types with various CPU, GPU, and memory options.
For the GPU-based dedispersion tasks (\textsc{cdmt} and \textsc{dedisp}), we use `A100' nodes, each of which have a 32-core Intel Xeon Gold 6326 CPU, \qty{1}{\tera\byte} of RAM, and dual NVIDIA A100 GPUs with \qty{80}{\giga\byte} of memory.
The CPU-based searching tasks (\textsc{accelsearch}, \textsc{accel\_sift.py}, and \textsc{prepfold}) are run on `KNL' nodes, each of which have a 68-core Intel Xeon Phi 7250 CPU and \qty{192}{\giga\byte} of RAM.
The \textsc{accelsearch} tasks are typically distributed over hundreds of KNL nodes to minimise the processing time.
Due to the large number of incoherent DM trials, writing the intermediate data products (time series, metadata, and candidate files) to the file system between processing steps can take a long time due to the large number of files.
We therefore group the intermediate data products into \textsc{tar} archives that are created and extracted in RAM on the processing nodes and written to the file system between processing steps.
For example, the \textsc{dedisp} task writes out archives containing 50 time series and their metadata, which are subsequently passed to \textsc{accelsearch} tasks where the extracted files are written directly to the RAM disk.
The typical run times for each processing task for an observation of length \qty{20}{\minute}s are summarised in Table~\ref{tab:benchmarks}.
The total pipeline run time is typically \qtyrange{3}{4}{\hour}, but can be longer depending on the availability of compute nodes.


\begin{table}[t]
    \centering
    \caption{Typical run times, number of tasks, and number of tasks per node for each of the processing steps in the critical path of the pipeline.}
    \label{tab:benchmarks}
    \begin{tabular}{llrrr}
        \toprule
        Task   & Node Type & Time/Task & \#Task & \#Task/Node \\
        \midrule
        \textsc{cdmt}           & A100  & 45\,min  & 1          & 1        \\
        \textsc{spp\_clean}     & KNL   & 10\,min  & 27         & 1--2  \\
        \textsc{dedisp}         & A100  & 5\,min   & 54         & 2        \\
        \textsc{accelsearch}    & KNL   & 30\,min  & 27000      & 50 \\
        \textsc{accel\_sift.py} & KNL   & 5\,min   & 1          & 1        \\
        \textsc{prepfold}       & KNL   & 50\,min  & $\sim$100  & $\sim$4  \\
        \bottomrule
    \end{tabular}
\end{table}

\begin{figure*}[p]
    \centering

    \includegraphics[width=0.49\linewidth]{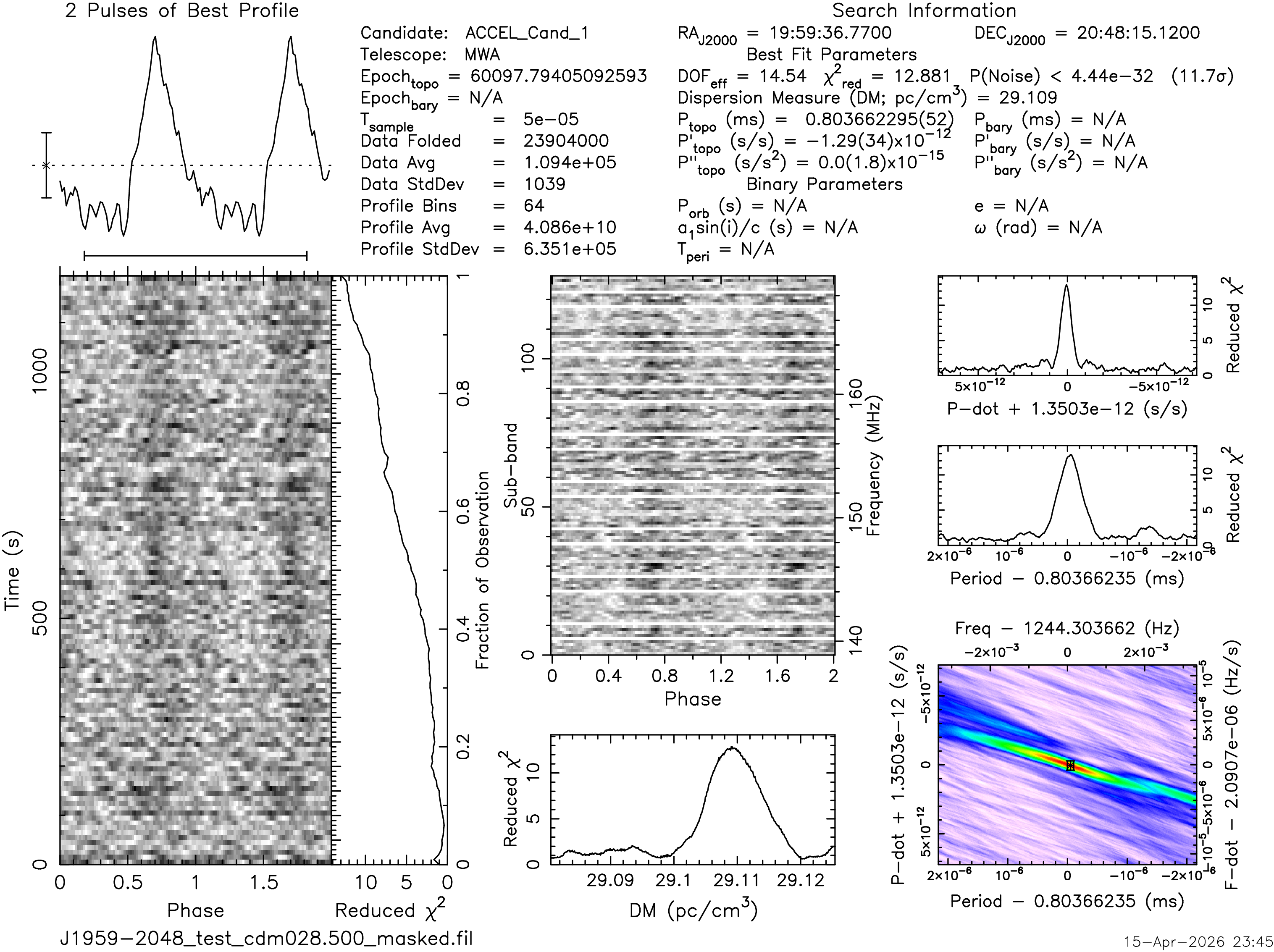}
    \includegraphics[width=0.49\linewidth]{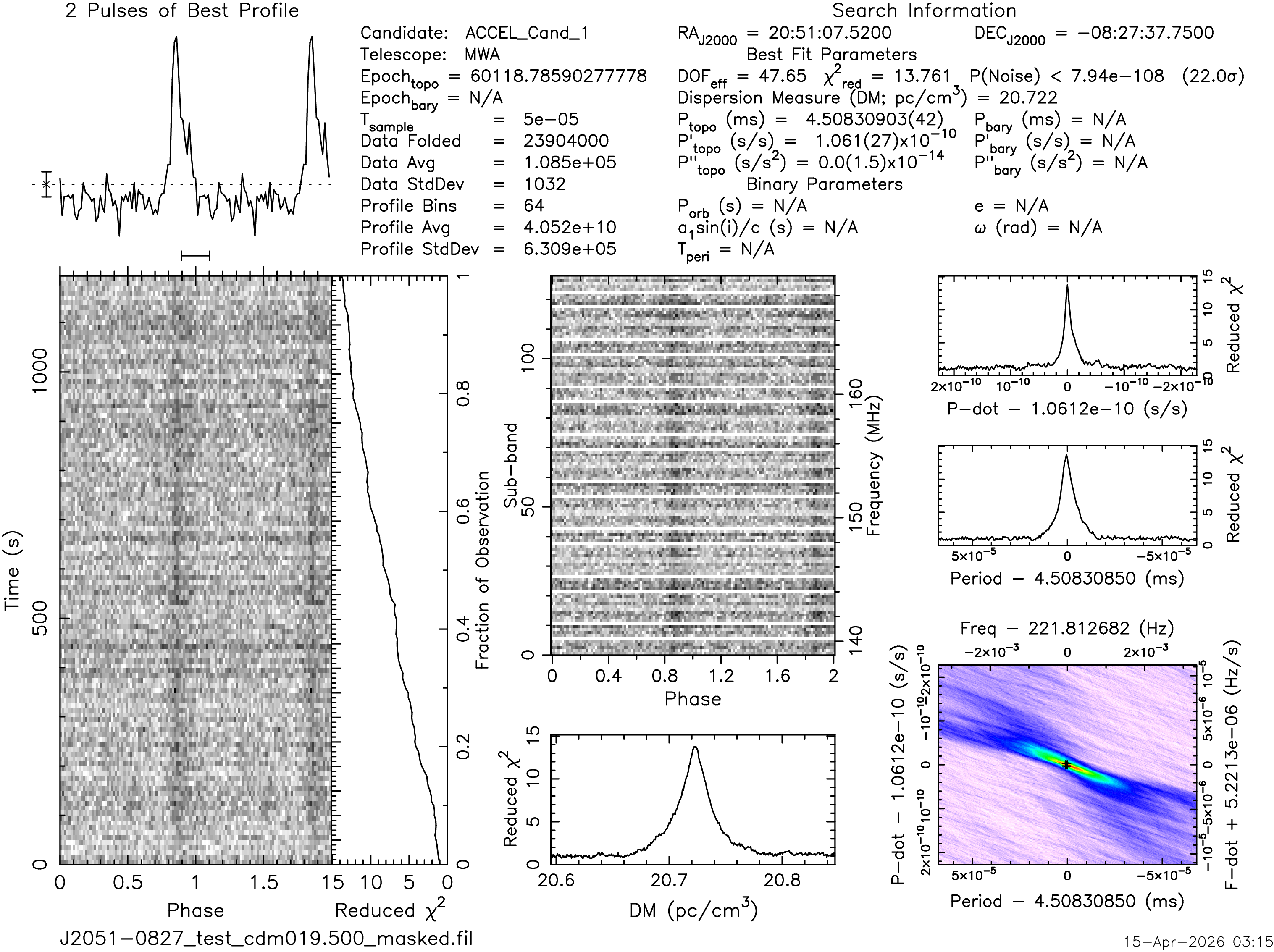}
    
    \vspace{2mm}
    \includegraphics[width=0.49\linewidth]{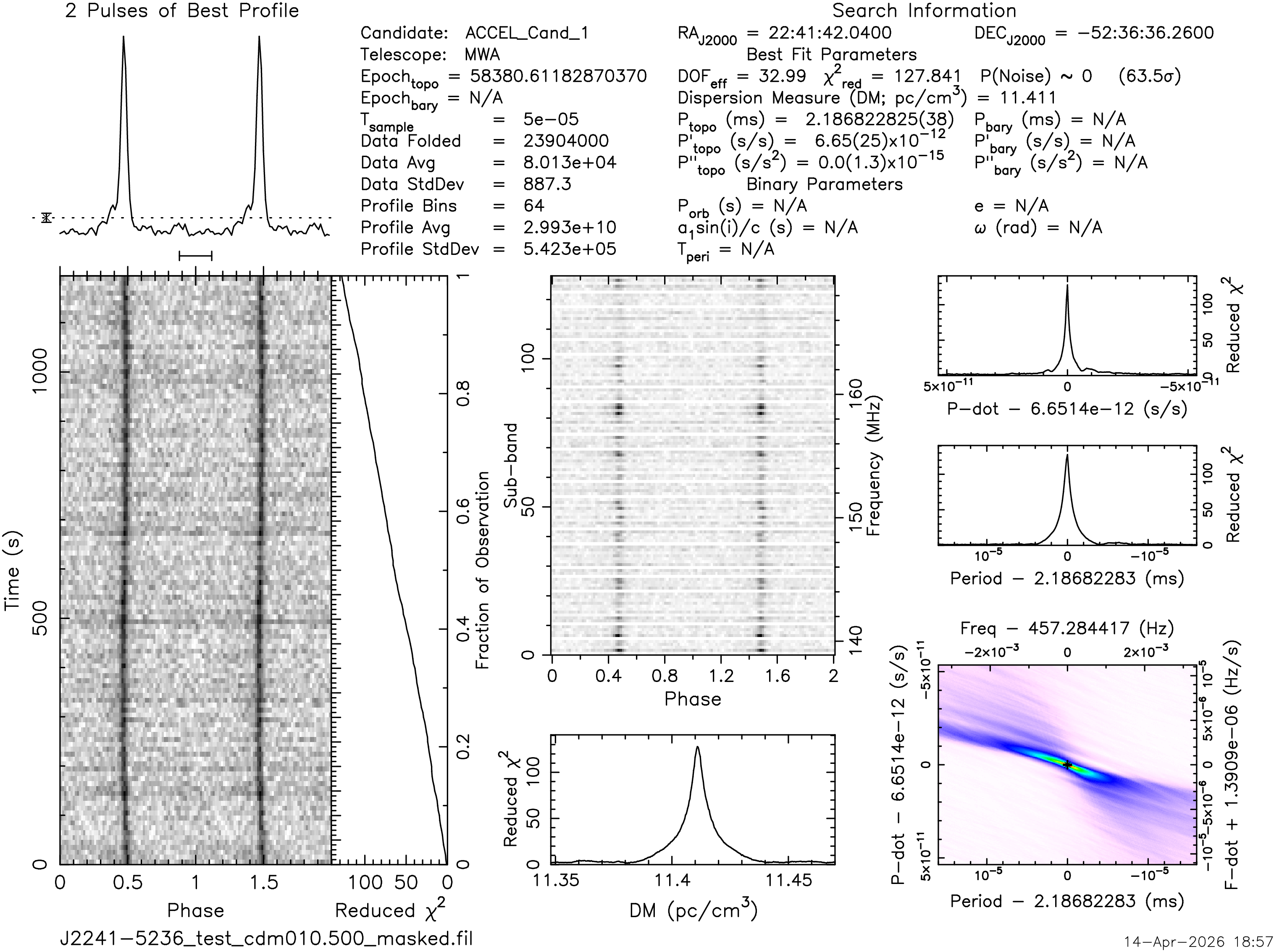}
    \includegraphics[width=0.49\linewidth]{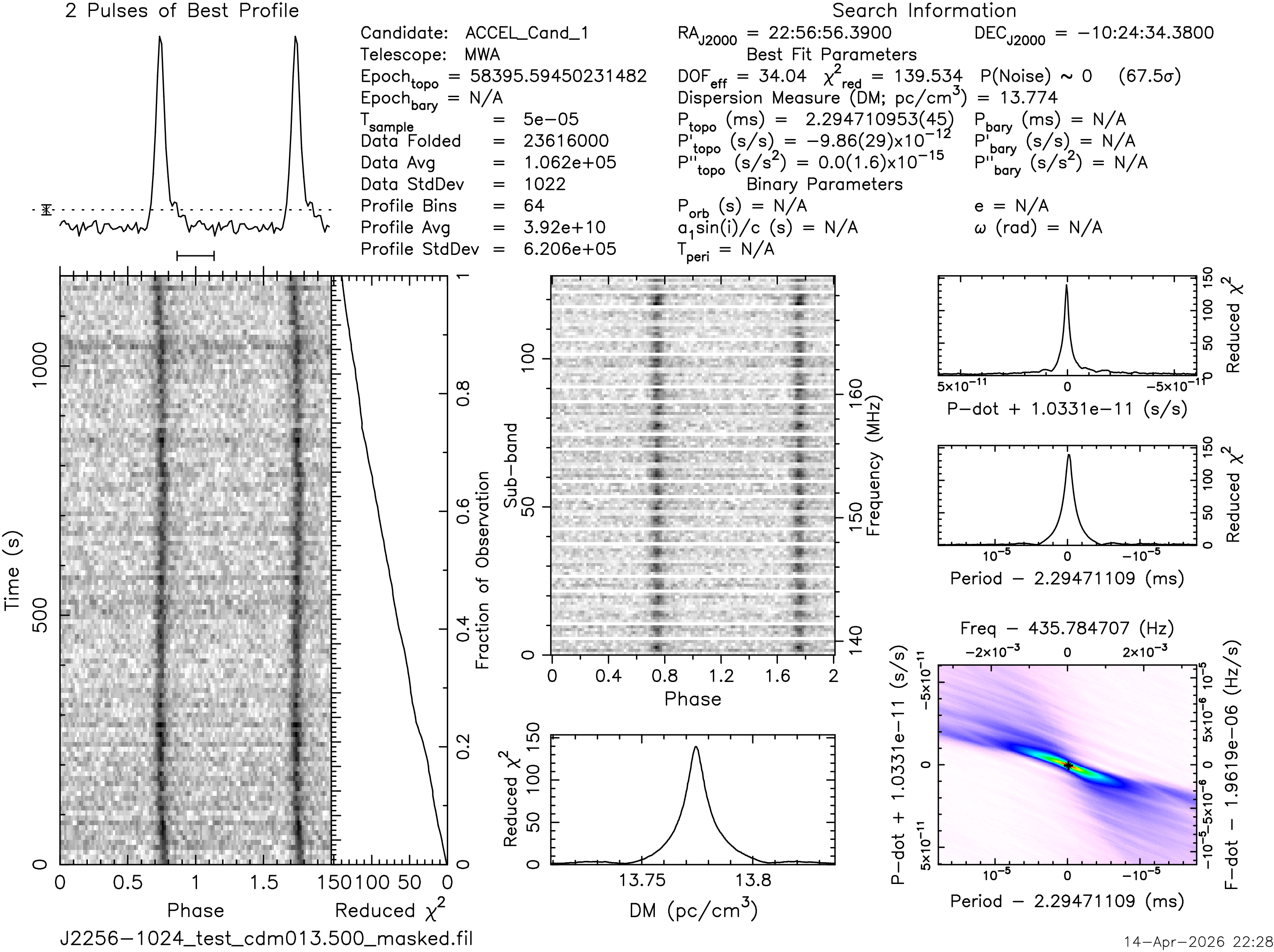}

    \vspace{2mm}
    \includegraphics[width=0.49\linewidth]{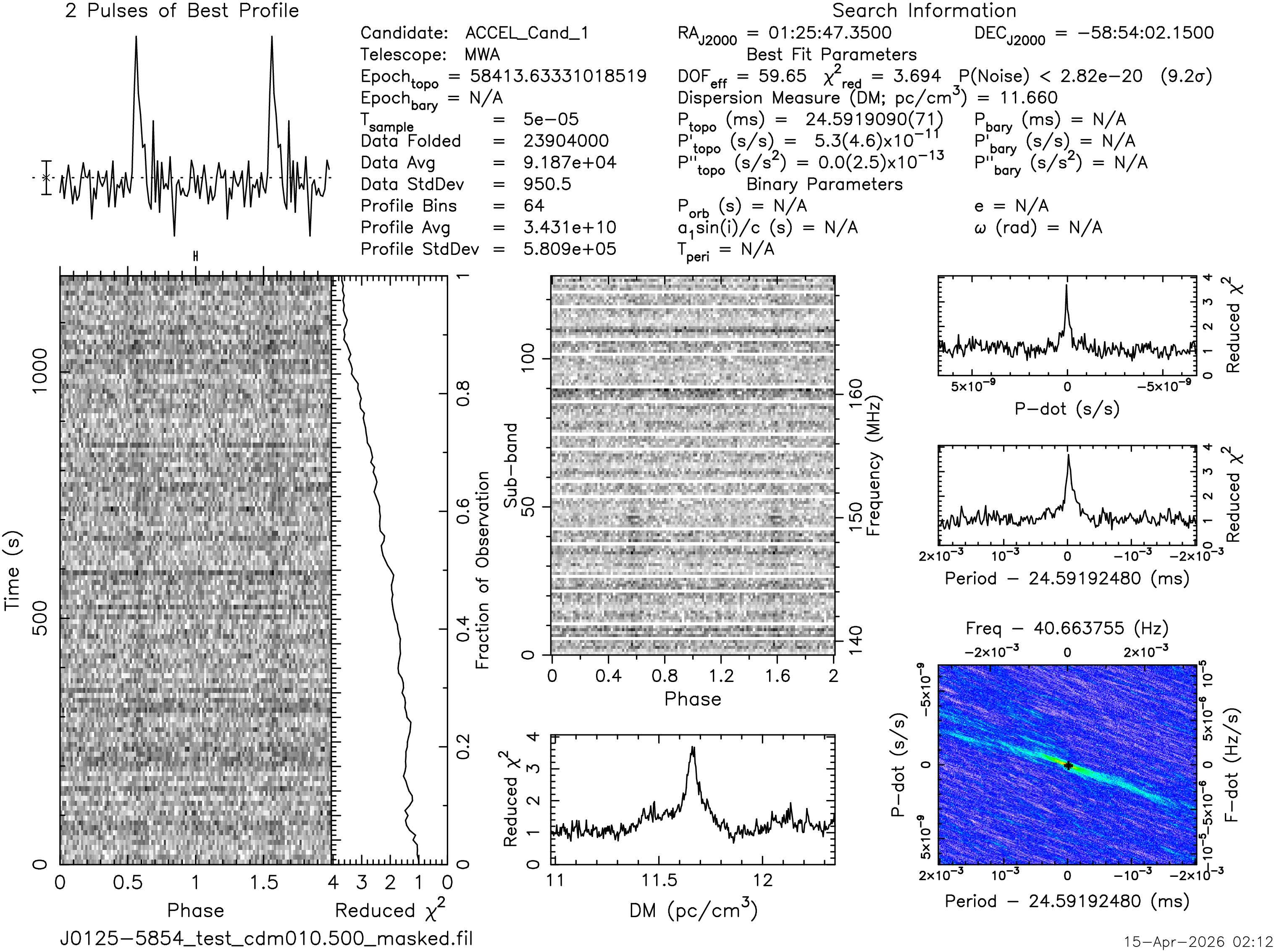}
    
    \caption{Candidate plots generated by \textsc{prepfold} for the five test pulsars blindly detected with the search pipeline: PSR J1959+2048 (top left), PSR J2051$-$0827 (top right), PSR J2241$-$5236 (centre left), PSR J2256$-$1024 (centre right), and PSR J0125$-$5854 (bottom). Each candidate plot shows the folded pulse profile over two full periods integrated over time and frequency (top left) and as a function of time and frequency (bottom left and centre). The reduced $\chi^2$ of the folded profile compared with noise is shown as a function of time and over the search ranges for the DM, $P$, $\dot{P}$, and the $P$--$\dot{P}$ plane. The horizontal bar below the integrated pulse profile shows the dispersive smearing if the observation were to be fully-incoherently dedispersed.}
    \label{fig:presto-plots}
\end{figure*}

\begin{table*}[t]
    \begin{threeparttable}
        \caption{Basic parameters and search results for the known pulsars used to test the pipeline. From left to right, the columns are: the pulsar's J-name, spin period ($P$), DM, and binary orbital period ($P_\mathrm{orb}$); the minimum smearing time given the true pulsar DM ($\tau_\mathrm{smear}^\mathrm{min}$); the smearing time for the candidate detected with the highest significance ($\tau_\mathrm{smear}^\mathrm{cand}$); the effective pulse duty cycle ($w_\mathrm{eff}$); the mean (i.e. period-averaged) flux density of the pulsar measured from the same observation ($S_\mathrm{mean}$); the estimated minimum detectable flux density at \qty{154}{\MHz} for the pulsar's spin period and duty cycle ($S_\mathrm{min}$); and the significances reported by \textsc{accelsearch} ($\sigma_\mathrm{FFT}$) and \textsc{prepfold} ($\sigma_\mathrm{fold}$). We report the statistical uncertainties for $S_\mathrm{mean}$ (the systematic uncertainties from the sensitivity simulations are omitted as they are correlated with $S_\mathrm{min}$).}
        \label{tab:known-psrs}
        \begin{tabular}{lcccccccccccccc}
            \toprule
            PSR & $P$ & DM & $P_\mathrm{orb}$ & $\tau_\mathrm{smear}^\mathrm{min}$ & $\tau_\mathrm{smear}^\mathrm{cand}$ & $w_\mathrm{eff}$ & $S_\mathrm{mean}$ & $S_\mathrm{min}$ & $\sigma_\mathrm{FFT}$ & $\sigma_\mathrm{fold}$ \\
            & [\unit{\ms}] & [\unit{\per\cm\cubed\pc}] & [\unit{\hour}] & [\unit{\ms}] & [\unit{\ms}] & & [\unit{\mJy}] & [\unit{\mJy}] & & \\
            \midrule
            J0125$-$5854$^\dagger$  & 24.592 & 11.665 & --   & 0.102 & 0.102 & 0.09 & 18(2)   & 15 & 9.74  & 9.2  \\
            J0952$-$0607$^\ddagger$ & 1.414  & 22.412 & 6.42 & 0.118 & --               & 0.34 & 56(5)   & 35 & --               & --   \\
            J1959+2048              & 1.607  & 29.108 & 9.17 & 0.103 & 0.345            & 0.26 & 260(12) & 74 & 12.10 & 11.7 \\
            J2051$-$0827            & 4.509  & 20.722 & 2.38 & 0.114 & 0.114 & 0.13 & 100(6)  & 24 & 22.51 & 22.0 \\
            J2241$-$5236            & 2.187  & 11.411 & 3.50 & 0.086 & 0.169            & 0.06 & 180(5)  & 13 & 66.85 & 63.5 \\
            J2256$-$1024            & 2.295  & 13.775 & 5.11 & 0.088 & 0.151 & 0.09 & 163(5)  & 16 & 68.25 & 67.5 \\
            \bottomrule
        \end{tabular}
        \begin{tablenotes}
            \item[$\dagger$] J0125$-$5854 is in a long binary orbit of order $\sim$years, so the acceleration is negligible.
            \item[$\ddagger$] J0952$-$0607 was not detected by the search pipeline in the test observation.
        \end{tablenotes}
    \end{threeparttable}
\end{table*}

\subsection{Testing}\label{sec:pipeline:testing}
Since a significant fraction of pulsars associated with \textit{Fermi}-LAT sources are spider pulsars, we tested the pipeline by targeting five known `black widow' pulsars that were detected in the SMART census of MSPs \citep{Lee2025}: PSRs J0952$-$0607, J1959+2048, J2051$-$0827, J2241$-$5236, and J2256$-$1024.
Notably, two of these pulsars (J0952$-$0607 and J2241$-$5236) were discovered in \textit{Fermi}-guided searches \citep{Bassa2017A&C,Keith2011}.
The short spin periods (\qtyrange{1.4}{4.5}{\ms}), short orbits (\qtyrange{2.4}{9.2}{\hour}), and range of pulse duty cycles (\qtyrange{6}{34}{\percent}) make these five pulsars realistic test cases.
We also targeted PSR J0125$-$5854, a partially-recycled binary pulsar in a long orbit that was discovered in the all-sky SMART pulsar survey (Tan et al. submitted).
For each pulsar, we formed a 20-min beam in the nearest SMART observation.
All pulsars were detectable when folded coherently with the timing ephemeris.
In Table~\ref{tab:known-psrs}, we list the basic pulsar parameters and the mean flux density measured using the method described in \citet{Lee2025}.
For comparison, we list the estimated minimum detectable flux density for a pulsar with the same spin period and duty cycle (see Section~\ref{sec:sensitivity} for details).

All of the test pulsars, besides J0952$-$0607, were blindly detected by the search pipeline and ranked as the top candidates in the search.
The non-detection of J0952$-$0607 appears to be due to the low signal-to-noise and/or the larger fractional smearing (i.e. $\tau_\mathrm{smear}^\mathrm{min}/P\approx\qty{8.3}{\percent}$), as different choices of the maximum $|z|$ and $N_\mathrm{harm}$ did not yield a detection.
J1959+2048 was detected at half the spin period (\qty{0.8}{\ms}) due to the pulse profile containing two similar pulses separated by $\sim$0.5 rotations.
As shown in Figure~\ref{fig:ddplan}, the second harmonic of J1959+2048 would be undetectable in a fully-incoherent search, so its detection provides validation that the pipeline is functioning as intended.
The \textsc{prepfold} plots of the five detected pulsars are shown in Figure~\ref{fig:presto-plots}.

As mentioned in Section~\ref{sec:pipeline:description}, when a candidate is detected in multiple DM trials, the sifting algorithm keeps only the trial with the highest $\sigma_\mathrm{FFT}$, which is subsequently folded.
Three of the test pulsars did not show the highest $\sigma_\mathrm{FFT}$ in the closest DM trial to the known pulsar DM.
In these cases, the best candidate has a DM step size smearing ($\tau_{\delta\mathrm{DM}}$) that is greater than optimal.
The minimum possible smearing time and the smearing time of the best candidate are listed in Table~\ref{tab:known-psrs}.

\section{Survey of Unassociated \textit{Fermi}-LAT Sources}\label{sec:survey}

\subsection{Target Selection and Survey Strategy}\label{sec:selection}

We created the target list by applying a set of selection criteria to the unassociated compact gamma-ray sources in the 4FGL-DR4 catalogue at declinations $<\qty{+25}{\degree}$ \citep{4FGL-DR4}.
We started by excluding sources for which the semi-major axis of the \qty{95}{\percent} confidence localisation ellipse ($r_{95}$) is greater than \qty{0.1}{\degree}.
As pulsars are typically steady sources of gamma-ray emission \citep{Kerr2025}, with the only exception being the state-changing gamma-ray pulsar PSR J2021+4026 \citep{Allafort2013}, we excluded sources with a 4FGL variability index greater than 28.
The gamma-ray spectral energy distributions (SEDs) of pulsars are typically modelled as a power-law with an exponential cutoff, peaking at around \qty{1.5}{\GeV} \citep[for a review of pulsar gamma-ray SEDs, see Section~6 of][]{3PC}.
We therefore also required that the gamma-ray SED shows at least $2\sigma$ of curvature compared with a simple power-law model, and the energy at the peak of the SED is less than \qty{5}{\GeV}.
Lastly, sources that were known to have been previously searched by the \textit{Fermi} Pulsar Search Consortium \citep{Ray2012}, the TRAPUM $L$-band survey \citep{Clark2023}, or Einstein@Home, were excluded.
Applying these selection criteria resulted in \num{271} pulsar candidates.
However, since many of the previously searched sources could be pulsars that were missed due to scintillation or radio eclipses, there is motivation to expand the candidate list to include these in future searches.
The constraint on gamma-ray variability could also be relaxed to search for more pulsars like J2021+4026.

In addition to our own target list, we also searched 4FGL-DR4 sources with radio associations in the GLEAM-X: Galactic Plane (GP) catalogue.
Tables 1 and 2 of \citet{Mantovanini2025} list the GLEAM-X: GP radio sources associated with 4FGL-DR4 sources, classified by whether or not they have previously been observed by Murriyang/Parkes as part of pulsar search campaigns.
Out of the 40 4FGL-DR4 sources that had not previously been searched, 37 were not in our initial target list, and were subsequently added.
We did not add any of the previously searched sources to our target list; however, 10 of these sources overlapped with our initial target list\footnote{The overlap is due to our list of previously-searched \textit{Fermi}-LAT sources being incomplete.}, so were searched anyway.

We note that some of the 4FGL-DR4 sources with GLEAM-X: GP associations have larger localisation uncertainties than the cutoff of our initial source selection: 24 sources have between $\qty{0.1}{\degree}<r_{95}<\qty{0.2}{\degree}$, and a further 8 sources have $r_{95}>\qty{0.2}{\degree}$.
The largest uncertainty is for 4FGL J1826.2$-$2830, with $r_{95}\approx\qty{0.305}{\degree}$.
The half-power tied-array beam width of the MWA Phase II Compact configuration is $\sim$\qty{22}{\arcminute} ($\sim$\qty{0.37}{\degree}) at zenith at \qty{154}{\MHz} \citep{Wayth2018,Meyers2026}.
For each source, we formed a single tied-array beam at the centre of the localisation ellipse.
The beam power (relative to the beam centre) for localisation errors of \qty{0.1}{\degree}, \qty{0.2}{\degree}, and \qty{0.3}{\degree} is approximately \qty{82}{\percent}, \qty{43}{\percent}, and \qty{11}{\percent}, respectively.

\begin{figure}[t]
    \centering
    \includegraphics[width=\linewidth]{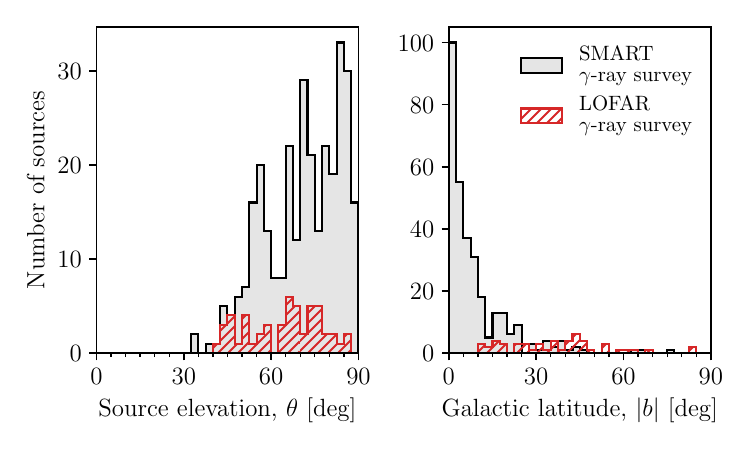}
    \caption{Distribution of source elevations (left) and Galactic latitudes (right) for the 308 \textit{Fermi}-LAT sources searched in this work (grey unhatched) and the 52 sources in the LOFAR gamma-ray survey \citep[red hatched;][]{Pleunis2017}.}
    \label{fig:samplehists}
\end{figure}

\begin{figure*}
    \centering
    \includegraphics[width=0.8\linewidth]{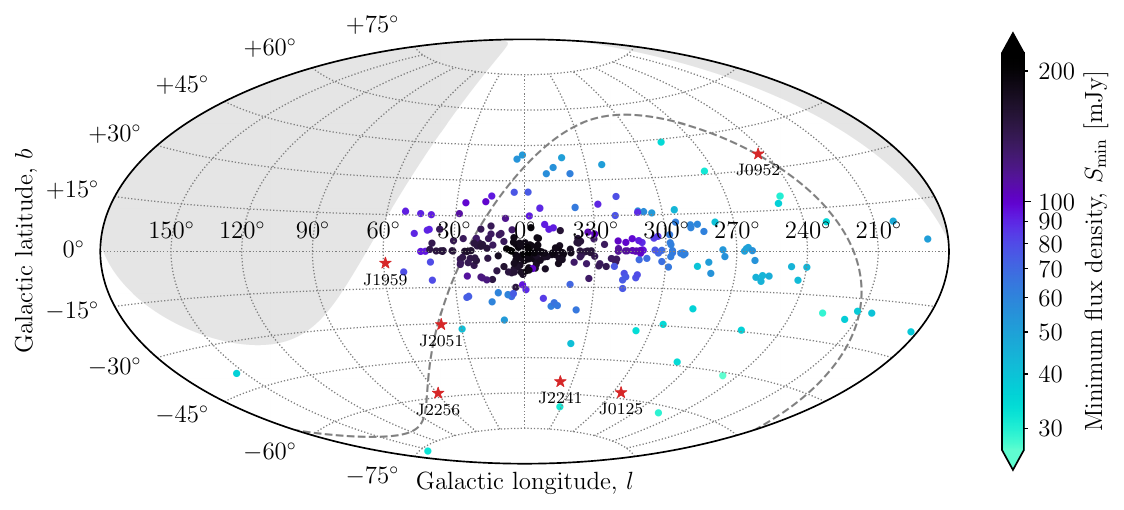}
    \caption{Galactic skymap of the 308 gamma-ray sources targeted in this survey (circles) and the six test pulsars (stars). The marker colours for the gamma-ray sources indicate the minimum detectable flux density ($S_\mathrm{min}$) for a spin period of \qty{2}{\ms}, a duty cycle of \qty{28}{\percent}, and an integration time of \qty{20}{\minute} in the beamformed SMART observations used in this work (assuming negligible scatter broadening). The grey shaded region indicates the declinations out of reach of the MWA. The grey dashed line shows the declination limit of the LOFAR gamma-ray survey \citep{Pleunis2017}.}
    \label{fig:skyplot}
\end{figure*}

The distributions of elevations and Galactic latitudes for the 308 targets in this work and the 52 targets from the LOFAR gamma-ray survey \citep{Pleunis2017} are shown in Figure~\ref{fig:samplehists}.
At high Galactic latitudes ($|b|>\qty{20}{\degree}$) there are a comparable number of targets from both surveys (36 for MWA vs. 40 for LOFAR).
At moderate latitudes ($\qty{10}{\degree}<|b|<\qty{20}{\degree}$) there are 49 MWA targets and 12 LOFAR targets.
The LOFAR survey excluded sources on the Galactic plane ($|b|<\qty{10}{\degree}$), whereas we include 223 targets at these latitudes.
Given that a large fraction of unidentified pulsars in the 4FGL catalogue are predicted to be on the Galactic plane \citep[e.g.][]{Sautron2025}, there is motivation to search at lower latitudes.
Additionally, the high density of unassociated gamma-ray sources on the Galactic plane reduces the processing overhead (e.g. transferring VCS data from the MWA archive to the cluster) per search compared with searches away from the plane.

Although each SMART observation is \qty{80}{\minute} in length, we chose to process \qty{20}{\minute} for each source for two reasons.
First, the smaller data size reduces the time required for data preparation and processing, including beamforming, file transfers, dedispersion, and searching.
This makes a larger and more uniform survey (in terms of source selection) more tractable.
Second, the shorter observation length improves sensitivity to shorter binary orbits.
Assuming that we are sensitive to binary pulsars with $P_\mathrm{orb}\gtrsim10T$ (in the worst-case orbital phase), we gain sensitivity to orbital periods between $\sim$\qty{3.3}{\hour} and $\sim$\qty{13.3}{\hour} by reducing the observation length from \qty{80}{\minute} to \qty{20}{\minute}.
This range includes $\sim$\qty{30}{\percent} of known radio-loud gamma-ray pulsars in binaries, with only $\sim$\qty{12}{\percent} having orbital periods less than \qty{3.3}{\hour} \citep{3PC}.
However, in future, deeper searches can be performed in the SMART data without the need for any additional observations.
For this survey, we have used the SMART observations for which each target was closest to the phase centre of the primary beam\footnote{For 4FGL J0524.7$-$8304, although it is closest to the phase centre of the southern celestial cap observation, we used the second-closest observation instead as it has a lower SEFD and better RFI conditions.}, and processed \qty{20}{\minute} of data during the period when the beam power is greatest towards the target position.

\subsection{Sensitivity}\label{sec:sensitivity}

To estimate the sensitivity of our survey, we first simulated the system temperature ($T_\mathrm{sys}$) and tied-array gain ($G$) using the method described by \citet{Meyers2017} and implemented by \citet{Lee2025}.
We then calculated the system equivalent flux density (SEFD) of the array as follows:
\begin{equation}\label{eq:sefd}
    \mathrm{SEFD} = f_\mathrm{c} \frac{T_\mathrm{sys}}{G},
\end{equation}
where $f_\mathrm{c}\geq1$ is a coherence factor that accounts for the inefficiencies in the tied-array beam formation that cause the real sensitivity to be less than ideal \citep[see Equation~2 of][]{Meyers2017}.
We assume a value of \num{1.43}, which is equivalent to a coherence `efficiency' of $1/f_\mathrm{c}\approx 0.70$.
At low frequencies, the system temperature is dominated by the sky temperature due to the steep power-law spectrum of the diffuse Galactic synchrotron emission.
We used the Haslam sky temperature map \citep{Haslam1982,Remazeilles2015} scaled to \qty{154.24}{\MHz} using a spectral index of \num{-2.55}, which is typical for the majority of the sky \citep{Guzman2011}.
Assuming that the average spectral index in the primary beam of an observation flattens to between \num{-2.3} and \num{-2.5} on the Galactic plane, we estimate that $T_\mathrm{sys}$ could be overestimated by between $\sim$\qtyrange{5}{30}{\percent}.
We calculated the SEFD at four frequencies and four time steps uniformly spaced over the frequency band and observing time, and report the mean of these 16 simulations in Table~\ref{tab:sources}.
The SEFD of our survey observations ranges between $\sim$\qtyrange{1}{8}{\kJy}, and is shown for each target on a Galactic sky map in Figure~\ref{fig:skyplot}.

We convert the SEFD into the minimum detectable flux density of a pulsar signal using the modified radiometer equation for pulsar observations \citep[Equation~A1.22 from][]{LorimerKramer2012}:
\begin{equation}\label{eq:smin}
    S_\mathrm{min} = \frac{\mathrm{S}/\mathrm{N}_\mathrm{min}\mathrm{SEFD}}{\sqrt{n_\mathrm{p}\Delta\nu \Delta t}} \sqrt{ \frac{W_\mathrm{eff}}{P - W_\mathrm{eff}} }
\end{equation}
where $\mathrm{S}/\mathrm{N}_\mathrm{min}\approx10$ is the signal-to-noise threshold, $n_\mathrm{p}=2$ is the number of instrumental polarisations, $\Delta\nu$ and $\Delta t$ are the observing bandwidth and integration time, $P$ is the spin period, and $W_\mathrm{eff}$ is the effective pulse width:
\begin{equation}
    W_\mathrm{eff} = \sqrt{(w_\mathrm{int} P)^2 + \tau_\mathrm{smear}^2 + \tau_\mathrm{scatt}^2}.
\end{equation}
Here, $w_\mathrm{int}$ is the intrinsic pulse width as a fraction of the pulse period (i.e. the pulse duty cycle), $\tau_\mathrm{smear}$ is the total smearing time from Equation~\eqref{eq:tsmear}, and $\tau_\mathrm{scatt}$ is the scattering time\footnote{As with the intrachannel smearing, we calculate the scattering time at the geometric centre frequency of the observing band (see footnote~\ref{fn:centrefreq}).}.
To estimate $\tau_\mathrm{scatt}$, we use the empirical model from \citet{Bhat2004}, which is a log-parabolic function of DM and a power-law function of observing frequency (with a spectral index of \num{-3.86}).
Based on the measurements from \citet{Bhat2004}, the DM limit imposed by scattering (where $P=W_\mathrm{eff}$) should be treated with around a factor of two uncertainty.
Using a sample of 878 non-recycled pulsars and 132 MSPs detected with MeerKAT, \citet{Karastergiou2024} found that pulse widths generally decrease with spin period following a power-law relation, $w_\mathrm{int}\propto P^{-0.308\pm 0.014}$.
We have used this model to estimate the typical duty cycle for a given spin period.

In Table~\ref{tab:sources}, we list $S_\mathrm{min}$ for each source at \qty{154.24}{\MHz}, calculated for a pulsar with a spin period of \qty{2}{\ms} and a duty cycle of \qty{28}{\percent}, where we assume the worst-case smearing of $\tau_\mathrm{smear}=\qty{136.1}{\us}$ and negligible scattering.
The $S_\mathrm{min}$ estimates range between $\sim$\qtyrange{30}{220}{\mJy} due to the different sky temperatures and offsets from the phase centre of the primary beam.
Assuming a spectral index of \num{-1.7} \citep[as used in 3PC;][]{3PC}, the equivalent $S_\mathrm{min}$ at \qty{1.4}{\GHz} is $\sim$\qtyrange{0.7}{5.2}{\mJy}.
These limits have an estimated systematic uncertainty of \qty{50}{\percent} due to the various assumptions made in the simulation.
This includes the coherence factor, the sky temperature (including the nominal value from Haslam and the assumed spectral index), and the beam model.

\begin{figure}[t]
    \centering
    \includegraphics[width=\linewidth]{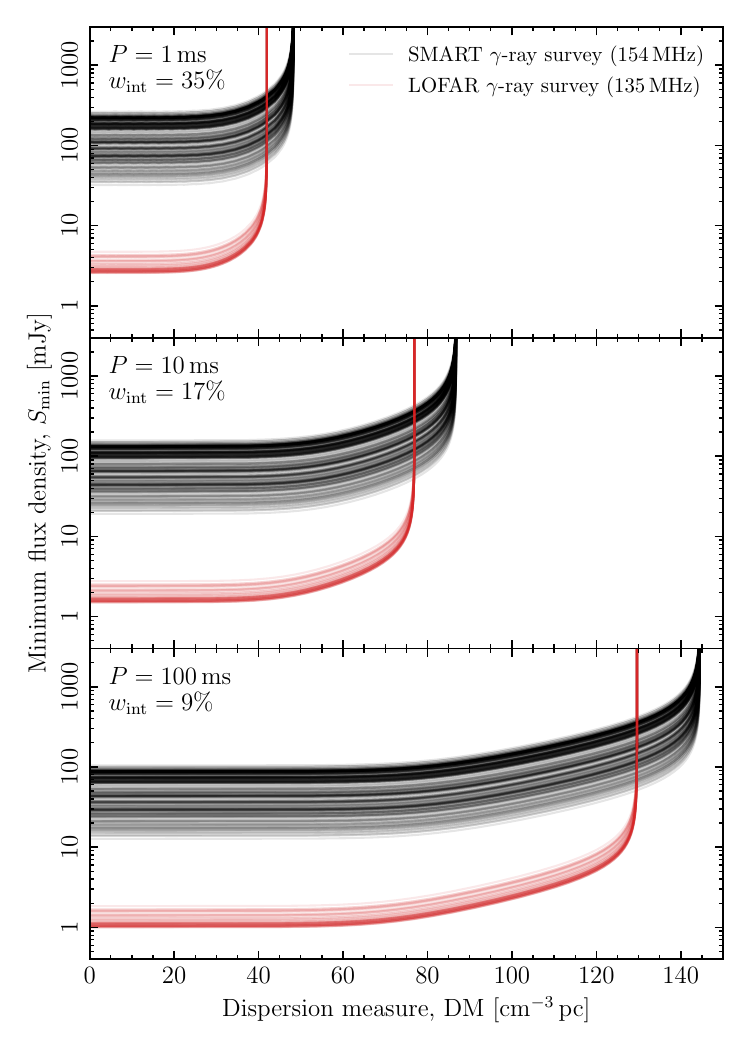}
    \caption{Minimum detectable pulsed flux density at \qty{154.24}{\MHz} as a function of DM for spin periods of 1, 10, and 100\,ms (from top to bottom). The assumed duty cycle for each spin period follows the power-law relation from \citet{Karastergiou2024}. The grey lines show the sensitivity curves for the 308 \textit{Fermi}-LAT sources searched in this work. The red lines show the sensitivity for the 52 sources in the LOFAR gamma-ray survey \citep{Pleunis2017}.}
    \label{fig:sensitivity}
\end{figure}

Figure~\ref{fig:sensitivity} shows how the survey sensitivity scales with DM for spin periods of \num{1}, \num{10}, and \qty{100}{\ms}.
Each subplot shows the sensitivity curves for all 308 targets in the survey.
For comparison, we also show the sensitivity of the LOFAR gamma-ray survey for the 52 targets listed in \citet{Pleunis2017}.
Following \citet{Pleunis2017}, we assume $G=\qty{5.6}{\K\per\Jy}$ for 21 LOFAR Core stations and $T_\mathrm{sys}=\qty{400}{\K}$ for all of the LOFAR observations.
Assuming the same coherence factor for LOFAR as for the MWA, we find $\mathrm{SEFD}\approx\qty{100}{\Jy}$.
Since the assumed gain for LOFAR only applies at zenith, we corrected for the sensitivity loss at lower elevations due to projection effects by scaling the sensitivity by $\sin^{-1.4}(\theta)$, where $\theta$ is the elevation angle \citep{Noutsos2015}.

The larger variance in the sensitivity of our survey is primarily due to the range of Galactic latitudes observed, and the per-source sensitivity simulations performed for the MWA targets.
At comparable Galactic latitudes and elevations (i.e. away from the Galactic plane), the simulated sensitivity of the MWA survey is around an order of magnitude lower than the LOFAR survey.
This is consistent with the expected difference due to the relative collecting areas and bandwidths of the telescopes.

\subsection{Search Results}

\begin{figure}[t]
    \centering
    \includegraphics[width=\linewidth]{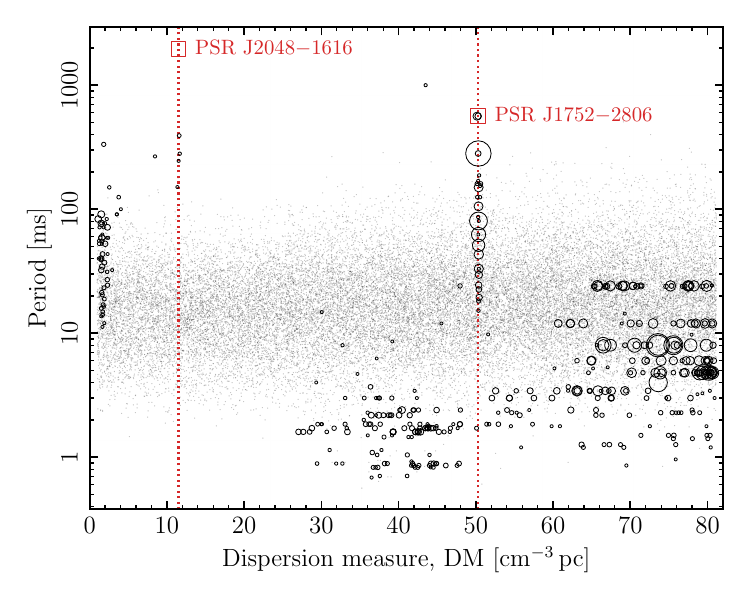}
    \caption{Distribution of independent pulsar candidates (i.e. after harmonic grouping per beam) identified across searches of 308 unidentified 4FGL sources. Candidates with $\sigma_\mathrm{FFT}>10$ are emphasised, with the marker size scaled by $\sigma_\mathrm{FFT}$. The two known pulsars identified in the searches are annotated at their spin period and DM, with a vertical line to show the harmonics associated with the pulsar.}
    \label{fig:cand-pdm}
\end{figure}

\begin{table}[t]
    \centering
    \caption{Summary of blind detections of known pulsars made in searches of 4FGL sources. For each pulsar, we list the J-name, spin period ($P$), and DM, followed by a list of pointings towards which the pulsar was detected. For each pointing, we list the MWA observation ID, the name of the 4FGL source that was being targeted, the offset between the known pulsar position and the phase centre of the tied-array beam, and the significance of the top candidate reported by \textsc{accelsearch} ($\sigma_\mathrm{FFT}$).}
    \label{tab:rediscovered-psrs}
    \begin{tabular}{lccc}
        \toprule
        Obs ID & 4FGL Name & Offset [\unit{\degree}] & $\sigma_\mathrm{FFT}$ \\
        \midrule
        \multicolumn{4}{c}{PSR J1752$-$2806 ($P=\qty{562.6}{\ms}$, $\mathrm{DM}=\qty{50.3}{\per\cm\cubed\pc}$)} \\
        1367946928 & J1714.8$-$1421 & 16.35 & 11.38 \\
        1367946928 & J1750.6$-$1906 & 9.02  & 19.88 \\
        1368640168 & J1752.3$-$2914 & 1.15  & 8.82 \\
        1367946928 & J1752.4$-$0758 & 20.13 & 9.17 \\
        1368640168 & J1753.2$-$2848 & 0.70  & 86.39 \\
        1368640168 & J1800.5$-$2910 & 1.96  & 8.72 \\
        1368640168 & J1828.2$-$3252 & 8.96  & 22.31 \\
        1368640168 & J1830.7$-$2414 & 9.31  & 27.46 \\
        \midrule
        \multicolumn{4}{c}{PSR J2048$-$1616 ($P=\qty{1961.6}{\ms}$, $\mathrm{DM}=\qty{11.5}{\per\cm\cubed\pc}$)} \\
        1371581520 & J2046.1$-$1626 & 0.60 & 13.17 \\
        \bottomrule
    \end{tabular}
\end{table}

In total, \num{31296} candidates passed the sifting criteria and were folded.
The distribution of these candidates in $P$--DM space is shown in Figure~\ref{fig:cand-pdm}.
The \num{417} candidates with $\sigma_\mathrm{FFT}>10$ are emphasised in the figure.
The clusters of candidates at DMs of \num{11.5} and \qty{50.3}{\per\cm\cubed\pc} are the harmonics of PSR J2048$-$1616 and PSR J1752$-$2806, which were blindly detected in sidelobes of the tied-array beam (see Table~\ref{tab:rediscovered-psrs} for a summary of these detections).
We also see the signatures of terrestrial RFI: the cluster of points near zero DM are due to broadband RFI, and the clusters in period spanning multiple DMs are due to narrowband RFI.
To ensure we did not miss strong candidates on first inspection, the harmonically-related candidates with $\sigma_\mathrm{FFT}>10$ were grouped together independent of DM to create a list of common RFI signals (i.e. `birdies').
The common RFI signals were then rejected, and the remaining candidates were carefully re-inspected.
No convincing pulsar candidates (besides known pulsars) were identified.

The tied-array beam of the MWA in the Phase II Compact configuration contains a regular pattern of sidelobes due to the two hexagonal tile clusters in the MWA core.
The main lobe is surrounded by six side lobes separated by $\sim$\qty{1}{\degree}, and there is a pattern of weaker grating lobes at separations of $\sim$\qty{9}{\degree} \citep[see Figure~7 of][]{Bhat2023a}.
The power in the side lobes depends on the primary beam, which has sensitivity to a significant fraction of the sky\footnote{Approximately \qty{40}{\percent} of the sky has a zenith-normalised primary beam power greater than \qty{1}{\percent} at \qty{154}{\MHz}.}.
It is therefore not surprising that J1752$-$2806, being one of the brightest pulsars in the southern sky ($S_\mathrm{mean}\sim\qty{3.6}{\Jy}$; Bhat et al. submitted), was detected as far as \qty{20}{\degree} away from the phase centre of the tied-array beam.
The regular tied-array beam pattern also explains why detections were made at approximately $(n\times \qty{9}{\degree})\pm\qty{2}{\degree}$ offsets (where $n=0, 1, 2$).
J2048$-$1616 was detected in a sidelobe at an offset of \qty{0.6}{\degree}, and is also a bright pulsar ($S_\mathrm{mean}\sim\qty{0.3}{\Jy}$; Bhat et al. submitted).

In Table~\ref{tab:sources}, we provide a list of the sources included in our survey, including flux density detection limits (see Section~\ref{sec:sensitivity} for details) and the maximum Galactic DM in the direction of the sources calculated using NE2025 \citep{Ocker2024,NE2025}.
The model indicates that we have searched up to the maximum Galactic DM for 34 sources.

\section{Discussion}\label{sec:discussion}

\begin{figure}[t]
    \centering
    \includegraphics[width=\linewidth]{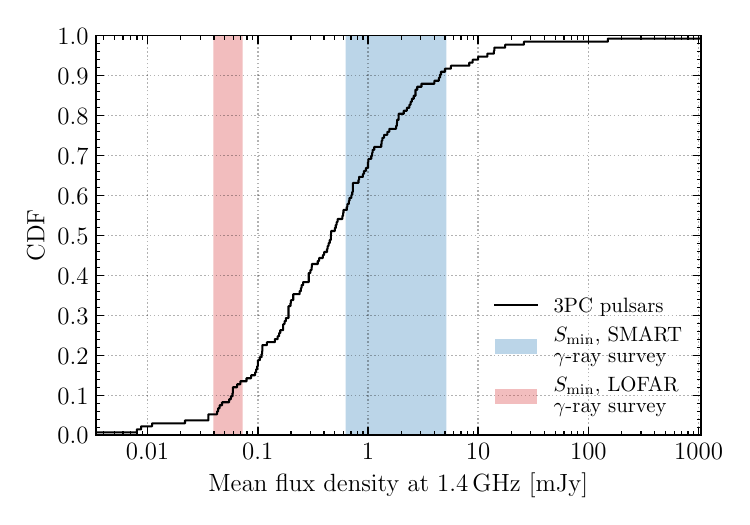}
    \caption{Cumulative distribution function (CDF) of the mean flux density at \qty{1.4}{\GHz} for the radio-loud gamma-ray pulsars in the 3PC catalogue with a DM less than \qty{100}{\per\cm\cubed\pc} \citep{3PC}. The shaded bands show the range of $S_\mathrm{min}$ for the SMART (blue) and LOFAR (red) gamma-ray surveys (see Section~\ref{sec:sensitivity}), scaled to \qty{1.4}{\GHz} assuming a spectral index of \num{-1.7}.}
    \label{fig:cdf}
\end{figure}

To better understand the null yield of our survey, we will compare our sensitivity to the properties of the known population of gamma-ray pulsars reported in the 3PC catalogue \citep{3PC}.
To mitigate any potential bias due to the completeness of the catalogue, we first restricted the sample to those with DMs less than \qty{100}{\per\cm\cubed\pc}.
Then, based on the sensitivity estimates in Section~\ref{sec:sensitivity} (for a spin period of \qty{2}{\ms}), and ignoring pulse broadening due to scattering, we find that our survey is sensitive to between \qtyrange{8}{44}{\percent} of gamma-ray pulsars at \qty{1.4}{\GHz}.
This is illustrated in Figure~\ref{fig:cdf}, showing the cumulative distribution of flux densities in the catalogue.
In contrast, the LOFAR gamma-ray survey was sensitive to \qtyrange{86}{95}{\percent} of gamma-ray pulsars, and discovered 3 MSPs from 52 unassociated sources \citep{Pleunis2017,Bassa2017ApJ,Bassa2018}.
Only one of those discoveries (J0952$-$0607) is expected to be bright enough to be detectable in the SMART data; although, as discussed in Section~\ref{sec:pipeline:testing}, it was not blindly detected by our pipeline in \qty{20}{\minute}, likely due to the low signal-to-noise.
If we compare only the sources searched at Galactic latitudes $|b|>\qty{10}{\degree}$, then based on the yield of the LOFAR survey, we should expect to discover $\sim$0.76 pulsars\footnote{The expected yield is the product of the number of LOFAR discoveries detectable by the MWA (1/52), the ratio of detectable samples (0.44/0.95), and the number of sources searched in the MWA survey at $|b|>\qty{10}{\degree}$ (85).}.
However, this assumes the best-case sensitivity, and it is also plausible that most of the undiscovered pulsars may be in the lower half of the flux-density distribution.
Therefore, this should be considered an upper limit on the number of expected discoveries.
Based on this estimate, the null yield of our survey is consistent with the yield of the LOFAR survey.

Given that the pilot survey we have presented here is limited by sensitivity, there are several ways in which we can improve the expected yield of future targeted surveys with the MWA.
First, the Phase III upgrade of the MWA enables observations with all 256 tiles in the Full Array configuration \citep{Tingay2026} -- double the number of tiles compared with the Phase II Compact configuration used for SMART.
This brings up to a factor-of-two improvement in the instantaneous sensitivity of the array.
Second, the real-time beamformer (RTB) under development for the MWA will greatly reduce the data volume and processing overhead compared with the VCS.
This will make it practical to plan targeted observations so that the phase centre of the primary beam is positioned closer to the source.
For sources located between SMART observations, beam offsets can be as large as \qty{10}{\degree} (see Table~\ref{tab:sources}), which can reduce the sensitivity by up to $\sim$\qty{50}{\percent}.
Therefore, in principle, the combination of the additional collecting area and targeted observations with the RTB will improve the instantaneous sensitivity by a factor of $\sim$\numrange{2}{3} compared with beamforming the SMART data.

Another consideration when planning future searches is the observing frequency.
Since the flux-density spectra of MSPs follow a steep power-law between \qtyrange{100}{200}{\MHz} \citep[i.e. they do not flatten like the spectra of non-recycled pulsars;][]{Kuzmin2001,Dowell2013,Kuniyoshi2015}, there can be a sensitivity advantage to observing at lower frequencies.
In order to estimate the relative sensitivities in different frequency bands, we first modelled the spectral dependence of the tied-array beam sensitivity.
We performed sensitivity simulations for all SMART pointings between \qtyrange{123.52}{215.68}{\MHz} in steps of \qty{15.36}{\MHz}, and fit power-law models to $T_\mathrm{sys}$ and $G$ as a function of frequency.
We noticed that the spectral dependence of $T_\mathrm{sys}$ deviates from a power-law at low elevations, so the pointings at elevations $\theta<\qty{50}{\degree}$ were excluded.
We also excluded pointings on the Galactic plane because our assumed spectral index for the sky temperature is most accurate at high Galactic latitudes.
For the remaining simulations, we used Equation~\eqref{eq:sefd} to estimate the spectral dependence of the SEFD from the best-fit models of $T_\mathrm{sys}$ and $G$.
We found $\mathrm{SEFD}\propto\nu^{-0.25\pm0.12}$, indicating that, on average, the array is slightly more sensitive at higher frequencies (within the considered frequency range).
Based on the spectra of pulsars discovered by LOFAR \citep{vanderWateren2023,Bassa2018}, we will assume that the pulsars most likely to be discovered with the MWA have spectra $S_\mathrm{mean}\propto\nu^{-2.5}$.
Therefore, the net sensitivity to these pulsars will on average scale as $S_\mathrm{mean}/\mathrm{SEFD}\propto\nu^{-2.25}$.

\begin{figure}[t]
    \centering
    \includegraphics[width=\linewidth]{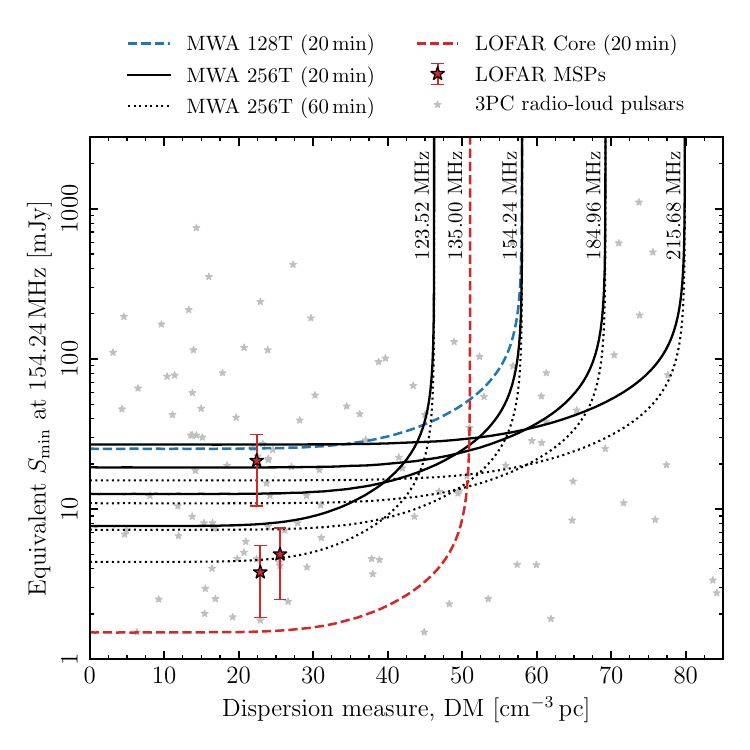}
    \caption{Equivalent minimum detectable flux density at \qty{154.24}{\MHz} as a function of DM for different telescope configurations and frequencies, assuming a spin period of \qty{2}{\ms} and a duty cycle of \qty{28}{\percent}. All sensitivity curves are scaled to \qty{154.24}{\MHz} assuming a spectral index of \num{-2.5}. The blue dashed line shows the MWA Phase II sensitivity (128 tiles) at \qty{154.24}{\MHz}, as used for the SMART survey. The black lines show the MWA Phase III Full Array sensitivity (256 tiles) at 4 centre frequencies. The red dashed line shows the LOFAR Core sensitivity at \qty{135}{\MHz}, as used in the targeted observations by \citet{Pleunis2017}. The integration times are 20 minutes for the dashed and solid lines, and 60 minutes for the dotted lines. The red stars indicate the reported mean flux densities at \qty{150}{\MHz} for the 3 MSPs discovered by the LOFAR gamma-ray survey: PSR J0653+4706 \citep{Bassa2018}, PSR J0952$-$0607 \citep{Bassa2017ApJ}, and PSR J1552+5437 \citep{Pleunis2017}. The grey stars show the mean flux densities of the radio-loud gamma-ray pulsars from the 3PC catalogue \citep{3PC}, scaled assuming a spectral index of \num{-1.7}.}
    \label{fig:compare-freqs}
\end{figure}

In Figure~\ref{fig:compare-freqs}, we compare the search sensitivity between the MWA Phase II Compact configuration (128 tiles) at \qty{154.24}{\MHz} and the MWA Phase III Full Array configuration (256 tiles) at 4 different frequency bands.
The tied-array sensitivity simulations discussed in Section~\ref{sec:sensitivity} indicate that the MWA Phase II Compact achieves a SEFD of $\sim$\qty{1}{\kJy} when observing away from the Galactic plane, at high elevations, and at small beam offsets (assuming $f_\mathrm{c}=1.43$).
We estimated the Phase III Full Array sensitivity by scaling the SEFD as follows:
\begin{equation}
    \mathrm{SEFD} \simeq 1\,\mathrm{kJy}\ \Biggl(\frac{N_\mathrm{T}}{128}\Biggr)^{-1}\Biggl(\frac{\nu_\mathrm{ctr}}{154.24\,\mathrm{MHz}}\Biggr)^{-0.25},
\end{equation}
where $N_\mathrm{T}$ is the number of tiles and $\nu_\mathrm{ctr}$ is the centre frequency of observation.
Figure~\ref{fig:compare-freqs} highlights a trade-off in sensitivity between lower and higher DMs.
At lower frequencies, there is an improvement in the maximum sensitivity at lower DMs due to pulsars being intrinsically brighter, but the increased scattering reduces sensitivity at higher DMs.
The opposite is true at higher frequencies: sensitivity is lost at lower DMs, but gained at higher DMs.
For example, the Phase III Full Array at \qty{215.68}{\MHz} would have a similar equivalent sensitivity to the Phase II Compact at \qty{154.24}{\MHz}, but the effective DM limit would be increased by $\sim$\qty{20}{\per\cm\cubed\pc} due to the reduced scattering.

\begin{table}[t]
    \centering
    \caption{Semi-coherent dedispersion plans for 4 different MWA frequency bands, for a total bandwidth of \qty{30.72}{\MHz} and a sample time of \qty{50}{\us}. From left to right, the columns are: the centre frequency of observation ($\nu_\mathrm{ctr}$), the coherent and incoherent DM trial step sizes ($\delta\mathrm{DM}_\mathrm{c,i}$), the maximum temporal smearing ($\tau_\mathrm{smear}$), and the total number of coherent and incoherent DM trials needed to search up to at least \qty{80}{\per\cm\cubed\pc} (\#$\mathrm{DM}_\mathrm{c,i}$).}
    \label{tab:all-ddplans}
    \begin{tabular}{lccccc}
        \toprule
        $\nu_\mathrm{ctr}$ & $\delta\mathrm{DM}_\mathrm{c}$ & $\delta\mathrm{DM}_\mathrm{i}$ & Max. $\tau_\mathrm{smear}$ & \#$\mathrm{DM}_\mathrm{c}$ & \#$\mathrm{DM}_\mathrm{i}$ \\
        $\text{[}$\unit{\MHz}] & [\unit{\per\cm\cubed\pc}] & [\unit{\per\cm\cubed\pc}] & [\unit{\us}] & & \\
        \midrule
        123.52 & 1.0 & 0.002 & 154.9 & 80 & 40000 \\
        154.24 & 3.0 & 0.003 & 136.1 & 27 & 27000 \\
        184.96 & 5.0 & 0.005 & 131.6 & 16 & 16000 \\
        215.68 & 5.0 & 0.010 & 143.9 & 16 & 8000 \\
        \bottomrule
    \end{tabular}
\end{table}

It is also important to consider how the computational cost scales with frequency.
For comparison, we chose dedispersion plans for each frequency band that keep the maximum smearing time to $\tau_\mathrm{smear}\lesssim\qty{150}{\us}$ (see Table~\ref{tab:all-ddplans}).
When moving down in frequency from \qty{154.24}{\MHz} to \qty{123.52}{\MHz}, the number of coherent DM trials approximately triples and the number of incoherent trials increases by $\sim$\qty{50}{\percent}.
Conversely, moving up in frequency to \qty{215.68}{\MHz} would reduce the number of coherent DM trials by $\sim$\qty{40}{\percent} and the number of incoherent DM trials by $\sim$\qty{70}{\percent}.
Additionally, since the Phase III Full Array beam is approximately $10\times$ smaller than the Phase II Compact beam, the localisation regions of unassociated \textit{Fermi}-LAT sources will need to be covered by multiple beams, increasing the computational cost per source compared with SMART.
For targeted searches, the computational cost of searches in all of these observing configurations are still tractable.

In the Phase III Full Array configuration at \qty{154.24}{\MHz}, we expect to be sensitive to up to $\sim$\qty{60}{\percent} of gamma-ray pulsars (below the DM limit imposed by scattering) in \qty{20}{\minute} observations.
This will increase the expected yield by $\sim$\qty{30}{\percent} relative to the pilot survey for the same number of sources searched.
However, it is also notable that the LOFAR discoveries lie in the lower end of the flux density distribution, with only one of the three MSPs being detectable by the Phase III Full Array in \qty{20}{\minute} (see Figure~\ref{fig:compare-freqs}).
This suggests that the instantaneous sensitivity may still be insufficient.
A reasonable option would be to increase the integration time at the sacrifice of sensitivity to short orbital periods.
Integrating for \qty{60}{\minute} instead of \qty{20}{\minute} lowers $S_\mathrm{min}$ by $\sim$\qty{40}{\percent}, but loses sensitivity to orbital periods $\lesssim\qty{10}{\hour}$.
Therefore, future targeted surveys with the RTB could maximise the discovery space by subdividing long observations and searching a range of time series lengths (e.g. \qty{20}{\minute}, \qty{40}{\minute}, and \qty{60}{\minute}).
In \qty{60}{\minute}, we expect to be sensitive to $\sim$\qty{70}{\percent} of gamma-ray pulsars at \qty{154.24}{\MHz} (below the DM limit imposed by scattering), increasing the expected yield by $\sim$\qty{60}{\percent} relative to the pilot survey.

The discovery potential could also be improved by revisiting sources multiple times.
A large fraction of gamma-ray MSPs are spider binaries, often exhibiting radio eclipses for a significant fraction of their orbit (i.e. minutes to hours) that increase in duration at low frequencies \citep[e.g.][]{Fruchter1988,Broderick2016,Polzin2020,Kumari2025}.
For typical spider eclipses of the order \qty{1}{\hour}, 2--3 visits of each source with observations of a similar length (i.e. $\sim$\qty{1}{\hour}) would be enough to mitigate most non-detections due to eclipses.
Furthermore, by separating the observations by weeks to months, one could also take advantage of potential flux density enhancements due to episodic (refractive) scintillation boosting.

\section{Conclusions and Future Work}\label{sec:conclusions}
We have developed a new pulsar search pipeline for the MWA, intended for targeted searches, which makes use of semi-coherent dedispersion to reduce dispersive smearing.
We used the GPU-accelerated \textsc{cdmt} algorithm originally developed for LOFAR to efficiently compute the coherent dedispersion trials.
To test the pipeline, we blindly detected five known binary MSPs.
Using the pipeline, we have performed the largest radio survey to date for pulsars towards unassociated \textit{Fermi}-LAT gamma-ray sources.
We selected a sample of 308 \textit{Fermi}-LAT sources from the 4FGL catalogue, and searched 20-min beamformed observations from the SMART survey data set.
No new pulsars were identified, although two known non-recycled pulsars were blindly detected in sidelobes of the tied-array beam.
We estimate flux density limits of between $\sim$\qtyrange{30}{220}{\mJy} at \qty{154.24}{\MHz} (or $\sim$\qtyrange{0.7}{5.2}{\mJy} at \qty{1.4}{\GHz}), for a spin period of \qty{2}{\ms} and a duty cycle of \qty{28}{\percent}.
This is sufficient to detect between \qtyrange{8}{44}{\percent} of radio-loud gamma-ray pulsars below the DM limit imposed by scattering.

The improved instantaneous sensitivity of the MWA Phase III, along with the development of a real-time beamformer, will enable future targeted searches with the MWA that are sensitive to $\sim$\qty{30}{\percent} more gamma-ray pulsars for the same integration time (\qty{20}{\minute}), or $\sim$\qty{60}{\percent} more pulsars for a \qty{1}{\hour} integration time.
To reach this sensitivity, sources on the Galactic plane (at $|b|<\qty{10}{\degree}$ and $\qty{330}{\degree}<l<\qty{30}{\degree}$) would be excluded.
Observing at a centre frequency of \qty{154.24}{\MHz} or lower is recommended to take advantage of the increase in signal-to-noise due to the intrinsic spectra of pulsars.
Sources should be revisited 2--3 times with observations of $\sim$\qty{1}{\hour} or longer separated by weeks to months in order to minimise non-detections due to spider pulsar eclipses or refractive scintillation.
Subdividing observations and searching time series of different lengths will help to maximise the sensitivity of acceleration searches given the limited instantaneous sensitivity of the array.

With a validated semi-coherent search pipeline, and the upgraded capability of the MWA Phase III, targeted pulsar searches with the MWA are now more practical than ever.
Besides \textit{Fermi}-LAT sources, targeted pulsar searches can also be performed on unidentified steep-spectrum radio sources detected in imaging surveys \citep[e.g.][]{Maan2018,Maan2026}, supernova remnants \citep[e.g.][]{Turner2024}, or globular clusters \citep[e.g.][]{Ridolfi2021,Das2025}.
Targeting candidates with known steep spectra will help to leverage the sensitivity improvement at low frequencies, and enable more competitive search sensitivities with the MWA.
The success of these pilot surveys will help to inform the potential significance of future SKA-Low surveys.

\begin{acknowledgement}

We thank Cees Bassa for providing advice on adapting \textsc{cdmt} for MWA data, and Chia Min Tan and Vivek Venkatraman Krishnan for helpful discussions on pulsar searching.
We also thank the anonymous reviewer for useful comments and suggestions that helped to improve the manuscript.
C.P.L. was supported by an Australian Government Research Training Program (RTP) Stipend and RTP Fee-Offset Scholarship (\url{https://doi.org/10.82133/C42F-K220}).
This scientific work made use of data obtained from Inyarrimanha Ilgari Bundara, the Murchison Radio-astronomy Observatory, operated by CSIRO.
We acknowledge the Wajarri Yamatji people as the traditional owners of the Observatory site.
This work was supported by resources provided by the Pawsey Supercomputing Research Centre’s Setonix Supercomputer (\url{https://doi.org/10.48569/18sb-8s43}), and their Acacia (\url{https://doi.org/10.48569/nfe9-a426}) and Banksia (\url{https://doi.org/10.48569/tnja-4s30}) Object Storage systems, with funding from the Australian Government and the Government of Western Australia.
This work made use of NASA's Astrophysics Data System and arXiv.

The \textit{Fermi} LAT Collaboration acknowledges generous ongoing support from a number of agencies and institutes that have supported both the development and the operation of the LAT as well as scientific data analysis. These include the National Aeronautics and Space Administration and the Department of Energy in the United States, the Commissariat \`a l'Energie Atomique and the Centre National de la Recherche Scientifique / Institut National de Physique Nucl\'eaire et de Physique des Particules in France, the Agenzia Spaziale Italiana and the Istituto Nazionale di Fisica Nucleare in Italy, the Ministry of Education, Culture, Sports, Science and Technology (MEXT), High Energy Accelerator Research Organization (KEK) and Japan Aerospace Exploration Agency (JAXA) in Japan, and the K.~A.~Wallenberg Foundation, the Swedish Research Council and the Swedish National Space Board in Sweden. Additional support for science analysis during the operations phase is gratefully acknowledged from the Istituto Nazionale di Astrofisica in Italy and the Centre National d'\'Etudes Spatiales in France. This work performed in part under DOE Contract DE-AC02-76SF00515.

Software/packages:
\textsc{cdmt} \citep{Bassa2017A&C},
\textsc{dedisp} \citep{Bassa2022},
\textsc{presto} \citep{Ransom2001},
\textsc{nextflow} \citep{DiTommaso2017},
\textsc{matplotlib} \citep{Matplotlib},
\textsc{numpy} \citep{NumPy},
\textsc{scipy} \citep{2020SciPy-NMeth},
\textsc{astropy} \citep{Astropy2013,Astropy2018,Astropy2022}.

\end{acknowledgement}
\printbibliography
\clearpage

\onecolumn

\begin{longtable}{lccccccccccl}
\caption{List of gamma-ray sources searched in this work. From left to right, the columns are: the source name from the 4FGL catalogue, the Galactic longitude ($l$) and latitude ($b$), the semi-major axis of the \qty{95}{\percent} confidence localisation ellipse ($r_{95}$), the epoch of the start of the observation, the MWA observation ID, the mean offset of the source from the phase centre of the primary beam, the source elevation ($\theta$), the system equivalent flux density (SEFD), the minimum detectable flux density at \qty{154.24}{\MHz} for a spin period of \qty{2}{\ms} and a duty cycle of \qty{28}{\percent} ($S_\mathrm{min}$), the maximum Galactic DM in the direction of the source from the NE2025 model \citep{NE2025}, and whether the source is associated with a radio source in the GLEAM-X: Galactic Plane catalogue \citep{Mantovanini2025}. Notes: $\dagger$ Sources previously targeted by Murriyang/Parkes pulsar searches \citep[see Section~4.1 and Table~2 of][]{Mantovanini2025}.} \label{tab:sources} \\

\toprule
\multicolumn{1}{l}{4FGL Name} &
\multicolumn{1}{c}{$l$} &
\multicolumn{1}{c}{$b$} &
\multicolumn{1}{c}{$r_{95}$} &
\multicolumn{1}{c}{Epoch} &
\multicolumn{1}{c}{Obs ID} &
\multicolumn{1}{c}{Offset} &
\multicolumn{1}{c}{$\theta$} &
\multicolumn{1}{c}{SEFD} &
\multicolumn{1}{c}{$S_\mathrm{min}$} &
\multicolumn{1}{c}{$\mathrm{DM}_\mathrm{max}$} &
\multicolumn{1}{l}{GLEAM-X} \\
\multicolumn{1}{l}{} &
\multicolumn{1}{c}{[\unit{\degree}]} &
\multicolumn{1}{c}{[\unit{\degree}]} &
\multicolumn{1}{c}{[\unit{\arcminute}]} &
\multicolumn{1}{c}{[MJD]} &
\multicolumn{1}{c}{} &
\multicolumn{1}{c}{[\unit{\degree}]} &
\multicolumn{1}{c}{[\unit{\degree}]} &
\multicolumn{1}{c}{[kJy]} &
\multicolumn{1}{c}{[mJy]} &
\multicolumn{1}{c}{[\unit{\per\cm\cubed\pc}]} &
\multicolumn{1}{l}{Assoc.} \\
\midrule
\endfirsthead

\multicolumn{12}{c}{Continued from previous page} \\
\toprule
\multicolumn{1}{l}{4FGL Name} &
\multicolumn{1}{c}{$l$} &
\multicolumn{1}{c}{$b$} &
\multicolumn{1}{c}{$r_{95}$} &
\multicolumn{1}{c}{Epoch} &
\multicolumn{1}{c}{Obs ID} &
\multicolumn{1}{c}{Offset} &
\multicolumn{1}{c}{$\theta$} &
\multicolumn{1}{c}{SEFD} &
\multicolumn{1}{c}{$S_\mathrm{min}$} &
\multicolumn{1}{c}{$\mathrm{DM}_\mathrm{max}$} &
\multicolumn{1}{l}{GLEAM-X} \\
\multicolumn{1}{l}{} &
\multicolumn{1}{c}{[\unit{\degree}]} &
\multicolumn{1}{c}{[\unit{\degree}]} &
\multicolumn{1}{c}{[\unit{\arcminute}]} &
\multicolumn{1}{c}{[MJD]} &
\multicolumn{1}{c}{} &
\multicolumn{1}{c}{[\unit{\degree}]} &
\multicolumn{1}{c}{[\unit{\degree}]} &
\multicolumn{1}{c}{[kJy]} &
\multicolumn{1}{c}{[mJy]} &
\multicolumn{1}{c}{[\unit{\per\cm\cubed\pc}]} &
\multicolumn{1}{l}{Assoc.} \\
\midrule
\endhead

\midrule
\multicolumn{12}{c}{Continued on next page} \\
\bottomrule
\endfoot

\bottomrule
\endlastfoot

J0122.4$-$1509 & 155.54 & $-$76.02 & 5.62 & 58427.603 & 1225462936 & 2.5 & 78.3 & 1.28 & 32 & 27 &  \\
J0230.3+1713 & 154.17 & $-$39.57 & 5.86 & 58423.651 & 1225118240 & 3.7 & 45.8 & 1.44 & 36 & 38 &  \\
J0243.1$-$4218 & 253.93 & $-$62.69 & 3.65 & 58420.660 & 1224859816 & 5.1 & 73.6 & 1.16 & 29 & 30 &  \\
J0347.0$-$6400 & 278.07 & $-$43.70 & 5.30 & 58778.759 & 1255803168 & 8.5 & 52.5 & 1.34 & 34 & 39 &  \\
J0414.7$-$4300 & 247.90 & $-$46.22 & 4.40 & 58757.818 & 1253991112 & 2.8 & 73.7 & 1.06 & 27 & 36 &  \\
J0433.0+1327 & 183.11 & $-$22.66 & 4.92 & 58764.809 & 1254594264 & 5.4 & 49.8 & 1.52 & 39 & 60 &  \\
J0524.7$-$8304 & 295.53 & $-$29.49 & 5.10 & 58778.801 & 1255803168 & 11.0 & 33.6 & 1.37 & 35 & 59 &  \\
J0529.9$-$0224 & 205.55 & $-$19.20 & 4.57 & 58785.794 & 1256407632 & 4.2 & 65.6 & 1.57 & 40 & 78 &  \\
J0536.1$-$1205 & 215.57 & $-$22.11 & 5.66 & 58820.705 & 1259427304 & 1.7 & 75.3 & 1.48 & 37 & 58 &  \\
J0541.4$-$0754 & 212.13 & $-$19.13 & 5.15 & 58820.707 & 1259427304 & 5.3 & 71.2 & 1.50 & 38 & 88 &  \\
J0553.9$-$5048 & 258.22 & $-$29.53 & 3.67 & 58799.772 & 1257617424 & 4.3 & 65.9 & 1.44 & 37 & 51 &  \\
J0558.1$-$2113 & 226.82 & $-$20.88 & 3.68 & 58792.778 & 1257010784 & 7.8 & 82.2 & 1.15 & 29 & 61 &  \\
J0618.9+2240 & 189.15 & +3.42 & 3.95 & 58823.724 & 1259685792 & 4.2 & 40.6 & 2.02 & 51 & 129 &  \\
J0705.0+1331 & 202.32 & +9.09 & 4.18 & 58823.739 & 1259685792 & 8.1 & 49.3 & 1.86 & 47 & 134 &  \\
J0708.8$-$3121 & 242.99 & $-$10.35 & 8.80 & 58841.710 & 1261241272 & 4.9 & 85.1 & 1.67 & 42 & 99 & Y \\
J0722.4$-$2650 & 240.26 & $-$5.67 & 4.10 & 58841.719 & 1261241272 & 1.1 & 88.9 & 1.72 & 44 & 154 & Y$^\dagger$ \\
J0735.0$-$7255 & 284.71 & $-$22.74 & 3.92 & 58907.574 & 1266932744 & 3.0 & 43.4 & 1.42 & 36 & 73 &  \\
J0736.9$-$3231 & 246.79 & $-$5.57 & 4.34 & 58841.729 & 1261241272 & 6.0 & 84.0 & 1.75 & 44 & 225 &  \\
J0739.6$-$4530 & 258.60 & $-$11.25 & 4.76 & 58896.588 & 1265983624 & 5.6 & 71.0 & 1.77 & 45 & 415 &  \\
J0741.9$-$4157 & 255.62 & $-$9.22 & 9.62 & 58896.588 & 1265983624 & 2.4 & 74.5 & 1.83 & 46 & 354 & Y \\
J0749.8$-$4420 & 258.44 & $-$9.10 & 5.62 & 58896.588 & 1265983624 & 4.1 & 72.3 & 1.74 & 44 & 449 &  \\
J0753.8$-$4700 & 261.13 & $-$9.81 & 8.18 & 58896.590 & 1265983624 & 6.7 & 69.6 & 1.79 & 45 & 508 & Y \\
J0800.7$-$1059 & 230.85 & +9.99 & 4.57 & 58898.590 & 1266155952 & 2.4 & 74.3 & 1.41 & 36 & 105 &  \\
J0828.4$-$4444 & 262.47 & $-$3.52 & 4.48 & 58896.614 & 1265983624 & 4.5 & 71.9 & 2.18 & 55 & 507 &  \\
J0848.8$-$4328 & 263.69 & +0.16 & 6.22 & 58896.629 & 1265983624 & 3.3 & 73.1 & 2.20 & 56 & 516 & Y \\
J0900.1$-$4402 & 265.45 & +1.36 & 5.50 & 58896.630 & 1265983624 & 4.2 & 72.4 & 1.98 & 50 & 453 &  \\
J0900.2$-$4608 & 267.03 & $-$0.01 & 16.34 & 58896.630 & 1265983624 & 6.1 & 70.4 & 2.01 & 51 & 480 & Y \\
J0900.5$-$4434 & 265.90 & +1.06 & 5.96 & 58896.630 & 1265983624 & 4.7 & 71.9 & 1.96 & 49 & 457 &  \\
J0901.1$-$4456 & 266.24 & +0.89 & 4.74 & 58896.630 & 1265983624 & 5.1 & 71.5 & 1.95 & 49 & 454 &  \\
J0906.8$-$2122 & 248.88 & +17.18 & 5.92 & 58890.658 & 1265470568 & 5.4 & 84.6 & 1.45 & 37 & 74 &  \\
J0910.1$-$1816 & 246.86 & +19.75 & 4.20 & 58898.625 & 1266155952 & 7.2 & 80.4 & 1.19 & 30 & 67 &  \\
J0919.5$-$6203 & 280.70 & $-$8.73 & 4.34 & 58900.634 & 1266329600 & 7.2 & 54.5 & 2.17 & 55 & 182 &  \\
J0926.1$-$5336 & 275.33 & $-$2.13 & 5.73 & 58900.634 & 1266329600 & 2.4 & 62.9 & 2.20 & 56 & 373 &  \\
J0942.1$-$5215 & 276.20 & +0.50 & 6.04 & 58900.634 & 1266329600 & 5.1 & 63.7 & 2.13 & 54 & 422 & Y \\
J0951.8$-$5944 & 282.03 & $-$4.38 & 5.02 & 58900.634 & 1266329600 & 7.1 & 56.2 & 2.03 & 51 & 319 &  \\
J1015.1$-$6353 & 286.73 & $-$6.05 & 4.79 & 58907.615 & 1266932744 & 9.9 & 51.4 & 1.56 & 39 & 280 &  \\
J1031.4$-$4448 & 278.42 & +11.31 & 4.41 & 58904.678 & 1266680784 & 4.6 & 71.8 & 1.59 & 40 & 135 &  \\
J1048.4$-$5030 & 283.78 & +7.73 & 2.00 & 58913.668 & 1267459328 & 4.5 & 66.0 & 2.41 & 61 & 214 & Y \\
J1105.1$-$2600 & 274.95 & +31.03 & 4.67 & 58923.651 & 1268321832 & 1.4 & 88.6 & 1.26 & 32 & 52 &  \\
J1109.1$-$4853 & 286.15 & +10.61 & 7.58 & 58913.679 & 1267459328 & 6.1 & 67.7 & 2.35 & 59 & 164 & Y \\
J1118.6$-$6159 & 292.27 & $-$1.06 & 4.50 & 58913.686 & 1267459328 & 7.1 & 54.6 & 2.46 & 62 & 666 &  \\
J1122.4$-$6237 & 292.90 & $-$1.51 & 4.69 & 58913.689 & 1267459328 & 7.8 & 53.9 & 2.41 & 61 & 602 & Y \\
J1123.2$-$5111 & 289.15 & +9.30 & 7.76 & 58913.689 & 1267459328 & 3.8 & 65.4 & 2.38 & 60 & 194 & Y \\
J1126.0$-$5007 & 289.20 & +10.46 & 4.01 & 58913.691 & 1267459328 & 4.9 & 66.4 & 2.34 & 59 & 173 &  \\
J1127.9$-$6158 & 293.30 & $-$0.68 & 6.04 & 58913.693 & 1267459328 & 7.1 & 54.6 & 2.39 & 61 & 712 & Y \\
J1133.7$-$6223 & 294.07 & $-$0.86 & 5.32 & 58913.697 & 1267459328 & 7.5 & 54.2 & 2.45 & 62 & 676 &  \\
J1203.5$-$1748 & 287.10 & +43.59 & 4.69 & 59299.665 & 1300809400 & 5.3 & 81.0 & 1.36 & 34 & 39 &  \\
J1203.7$-$6303 & 297.56 & $-$0.68 & 4.49 & 59304.651 & 1301240224 & 8.9 & 53.5 & 2.61 & 66 & 726 &  \\
J1204.5$-$5032 & 295.36 & +11.64 & 4.40 & 58913.710 & 1267459328 & 4.8 & 65.9 & 2.45 & 62 & 164 &  \\
J1205.3$-$6212 & 297.59 & +0.18 & 3.84 & 58913.710 & 1267459328 & 7.5 & 54.3 & 2.50 & 63 & 706 &  \\
J1207.4$-$4536 & 294.94 & +16.58 & 3.79 & 59306.645 & 1301412552 & 5.4 & 70.9 & 1.73 & 44 & 111 &  \\
J1207.9$-$6443 & 298.31 & $-$2.25 & 4.70 & 59304.651 & 1301240224 & 7.2 & 51.8 & 2.47 & 62 & 467 &  \\
J1208.0$-$6900 & 299.04 & $-$6.46 & 3.23 & 59304.651 & 1301240224 & 3.0 & 47.6 & 2.54 & 64 & 275 & Y$^\dagger$ \\
J1213.6$-$5954 & 298.20 & +2.62 & 5.50 & 58913.710 & 1267459328 & 5.8 & 56.5 & 2.43 & 61 & 435 &  \\
J1216.7$-$5911 & 298.49 & +3.38 & 5.37 & 58913.710 & 1267459328 & 5.4 & 57.1 & 2.47 & 62 & 385 &  \\
J1220.1$-$5558 & 298.53 & +6.64 & 4.53 & 58913.710 & 1267459328 & 4.0 & 60.2 & 2.51 & 64 & 269 &  \\
J1241.9$-$6706 & 302.01 & $-$4.25 & 3.50 & 59304.674 & 1301240224 & 4.9 & 49.5 & 2.68 & 68 & 345 &  \\
J1244.3$-$6233 & 302.11 & +0.31 & 2.83 & 59304.675 & 1301240224 & 9.4 & 54.0 & 2.71 & 69 & 721 &  \\
J1256.1$-$5652 & 303.57 & +5.99 & 5.75 & 59314.679 & 1302106648 & 5.0 & 59.1 & 3.40 & 86 & 293 &  \\
J1257.0$-$6339 & 303.55 & $-$0.79 & 2.37 & 59304.684 & 1301240224 & 8.3 & 52.9 & 2.79 & 71 & 778 &  \\
J1303.1$-$4714 & 305.00 & +15.59 & 4.74 & 59306.683 & 1301412552 & 7.0 & 69.3 & 2.09 & 53 & 128 &  \\
J1308.9$-$5730 & 305.29 & +5.29 & 4.64 & 59314.679 & 1302106648 & 3.8 & 58.9 & 3.30 & 83 & 316 &  \\
J1309.1$-$6223 & 304.98 & +0.41 & 4.72 & 59304.693 & 1301240224 & 9.6 & 54.2 & 3.03 & 77 & 946 &  \\
J1322.0$-$4624 & 308.42 & +16.13 & 4.16 & 59306.687 & 1301412552 & 6.6 & 70.0 & 2.27 & 57 & 126 &  \\
J1335.5$-$4546 & 310.92 & +16.41 & 5.05 & 59306.687 & 1301412552 & 7.3 & 70.1 & 2.23 & 56 & 125 &  \\
J1336.9$-$4611 & 311.09 & +15.95 & 5.85 & 59314.684 & 1302106648 & 8.8 & 70.4 & 3.49 & 88 & 129 &  \\
J1344.2$-$6517 & 308.42 & $-$2.99 & 3.78 & 59304.693 & 1301240224 & 7.4 & 50.9 & 2.72 & 69 & 424 &  \\
J1349.1$-$5829 & 310.43 & +3.54 & 5.60 & 59314.693 & 1302106648 & 3.7 & 58.1 & 3.51 & 89 & 398 & Y \\
J1351.6$-$6142 & 310.00 & +0.34 & 3.70 & 59314.695 & 1302106648 & 6.8 & 54.9 & 3.74 & 95 & 909 &  \\
J1353.5$-$6128 & 310.27 & +0.52 & 3.98 & 59314.696 & 1302106648 & 6.6 & 55.1 & 3.78 & 96 & 877 &  \\
J1357.8$-$5724 & 311.81 & +4.32 & 5.00 & 59314.699 & 1302106648 & 2.6 & 59.2 & 3.66 & 93 & 367 &  \\
J1400.0$-$2415 & 322.35 & +36.00 & 3.08 & 59309.714 & 1301674968 & 2.7 & 87.3 & 2.02 & 51 & 49 &  \\
J1400.5$-$6147 & 311.00 & $-$0.01 & 5.29 & 59314.701 & 1302106648 & 6.9 & 54.8 & 4.02 & 102 & 978 &  \\
J1401.9$-$6130 & 311.24 & +0.23 & 4.26 & 59314.702 & 1302106648 & 6.6 & 55.1 & 4.07 & 103 & 967 &  \\
J1403.5$-$6236 & 311.12 & $-$0.88 & 2.60 & 59314.703 & 1302106648 & 7.7 & 54.0 & 3.97 & 100 & 812 &  \\
J1408.7$-$3747 & 319.39 & +22.60 & 4.43 & 59316.709 & 1302282040 & 3.5 & 78.5 & 3.09 & 78 & 92 &  \\
J1415.2$-$5550 & 314.57 & +5.13 & 8.59 & 59314.711 & 1302106648 & 1.2 & 60.7 & 3.87 & 98 & 343 & Y \\
J1415.3$-$6110 & 312.87 & +0.07 & 3.43 & 59314.711 & 1302106648 & 6.3 & 55.4 & 4.27 & 108 & 1002 &  \\
J1415.4$-$6458 & 311.67 & $-$3.55 & 3.53 & 59304.693 & 1301240224 & 9.2 & 50.2 & 2.76 & 70 & 403 &  \\
J1418.7$-$6110 & 313.26 & $-$0.07 & 3.47 & 59314.714 & 1302106648 & 6.3 & 55.4 & 4.22 & 107 & 1003 &  \\
J1428.0$-$6026 & 314.58 & +0.23 & 4.42 & 59314.720 & 1302106648 & 5.6 & 56.2 & 4.19 & 106 & 1000 &  \\
J1431.0$-$4432 & 321.01 & +14.81 & 3.08 & 59316.717 & 1302282040 & 4.3 & 72.0 & 3.05 & 77 & 147 &  \\
J1437.6$-$5616 & 317.33 & +3.60 & 11.36 & 59314.721 & 1302106648 & 1.9 & 60.3 & 3.97 & 100 & 419 & Y \\
J1443.7$-$7037 & 312.05 & $-$9.77 & 4.66 & 60069.723 & 1367342464 & 6.8 & 44.4 & 3.14 & 79 & 216 & Y \\
J1444.9$-$5939 & 316.83 & +0.11 & 4.81 & 59314.721 & 1302106648 & 5.3 & 56.8 & 4.49 & 114 & 1040 &  \\
J1449.8$-$5923 & 317.50 & +0.09 & 5.49 & 59314.721 & 1302106648 & 5.3 & 57.0 & 4.51 & 114 & 1053 &  \\
J1450.8$-$1424 & 341.51 & +39.36 & 4.47 & 59321.716 & 1302712864 & 2.0 & 77.7 & 2.00 & 51 & 43 &  \\
J1451.6$-$3726 & 327.88 & +19.54 & 4.97 & 59316.731 & 1302282040 & 3.1 & 79.1 & 3.56 & 90 & 107 &  \\
J1456.3$-$5419 & 320.58 & +4.21 & 5.18 & 59314.721 & 1302106648 & 4.0 & 61.8 & 4.02 & 102 & 398 &  \\
J1456.4$-$7129 & 312.59 & $-$11.00 & 3.93 & 60069.723 & 1367342464 & 5.6 & 44.0 & 3.06 & 78 & 194 &  \\
J1458.8$-$2120 & 338.60 & +32.57 & 2.74 & 60113.573 & 1371131072 & 8.1 & 81.9 & 2.44 & 62 & 55 &  \\
J1506.5$-$5708 & 320.49 & +1.05 & 4.27 & 59314.721 & 1302106648 & 5.7 & 58.7 & 4.40 & 111 & 809 & Y$^\dagger$ \\
J1511.2$-$5803 & 320.57 & $-$0.06 & 4.69 & 59314.721 & 1302106648 & 6.6 & 57.7 & 4.51 & 114 & 1164 &  \\
J1513.7$-$1519 & 346.36 & +35.27 & 3.29 & 59321.732 & 1302712864 & 2.8 & 78.6 & 2.18 & 55 & 50 &  \\
J1529.4$-$6027 & 321.24 & $-$3.33 & 12.81 & 60070.720 & 1367428632 & 7.6 & 55.5 & 5.35 & 135 & 452 & Y \\
J1530.0$-$1522 & 349.89 & +32.68 & 4.97 & 59321.737 & 1302712864 & 3.5 & 78.5 & 2.24 & 57 & 54 &  \\
J1532.9$-$0313 & 1.39 & +40.64 & 5.26 & 60071.692 & 1367512184 & 5.1 & 66.6 & 2.09 & 53 & 41 &  \\
J1534.0$-$5232 & 326.29 & +2.79 & 3.33 & 60070.720 & 1367428632 & 5.7 & 63.2 & 5.39 & 136 & 483 & Y$^\dagger$ \\
J1534.3$-$3312 & 338.04 & +18.34 & 3.97 & 60113.579 & 1371131072 & 6.7 & 83.3 & 2.51 & 64 & 109 &  \\
J1534.4$-$6719 & 317.71 & $-$9.25 & 5.90 & 60069.723 & 1367342464 & 5.5 & 49.0 & 3.09 & 78 & 233 &  \\
J1536.5$-$6855 & 316.90 & $-$10.66 & 4.59 & 60069.723 & 1367342464 & 4.0 & 47.4 & 3.00 & 76 & 204 &  \\
J1537.3$-$6110 & 321.61 & $-$4.47 & 8.57 & 60070.720 & 1367428632 & 7.5 & 55.0 & 5.41 & 137 & 389 & Y \\
J1538.0$-$4638 & 330.32 & +7.18 & 5.38 & 59316.751 & 1302282040 & 7.1 & 69.6 & 4.00 & 101 & 281 &  \\
J1542.1$-$5901 & 323.38 & $-$3.11 & 5.40 & 60070.720 & 1367428632 & 5.5 & 57.3 & 5.26 & 133 & 477 &  \\
J1543.6$-$0244 & 4.09 & +38.89 & 3.37 & 60071.700 & 1367512184 & 4.7 & 66.1 & 2.16 & 55 & 43 &  \\
J1547.4$-$4802 & 330.72 & +5.11 & 6.49 & 60070.720 & 1367428632 & 7.7 & 68.2 & 5.23 & 132 & 316 & Y \\
J1548.1$-$4416 & 333.18 & +7.97 & 3.88 & 59316.751 & 1302282040 & 6.5 & 71.4 & 4.24 & 107 & 246 &  \\
J1550.3$-$6223 & 322.09 & $-$6.38 & 9.54 & 60069.723 & 1367342464 & 9.8 & 54.1 & 2.99 & 76 & 317 & Y \\
J1556.4$-$6911 & 318.14 & $-$11.98 & 5.07 & 60069.723 & 1367342464 & 3.0 & 47.4 & 2.96 & 75 & 183 &  \\
J1611.6$-$6013 & 325.46 & $-$6.43 & 5.38 & 60070.722 & 1367428632 & 5.3 & 56.4 & 5.66 & 143 & 321 &  \\
J1613.0$-$5102 & 331.91 & +0.11 & 3.40 & 60070.723 & 1367428632 & 4.0 & 65.6 & 6.19 & 157 & 1224 &  \\
J1615.3$-$1512 & 358.67 & +24.91 & 5.84 & 60076.719 & 1367946928 & 4.6 & 77.9 & 2.97 & 75 & 77 &  \\
J1615.9$-$4322 & 337.58 & +5.33 & 5.00 & 60083.700 & 1368550080 & 4.2 & 73.0 & 6.22 & 157 & 379 &  \\
J1616.0$-$4501 & 336.43 & +4.13 & 10.14 & 60083.700 & 1368550080 & 5.5 & 71.3 & 6.19 & 157 & 420 & Y \\
J1616.6$-$5341 & 330.48 & $-$2.17 & 2.72 & 60070.726 & 1367428632 & 1.5 & 62.9 & 5.63 & 143 & 575 &  \\
J1618.0$-$5119 & 332.28 & $-$0.63 & 5.96 & 60070.727 & 1367428632 & 3.7 & 65.3 & 6.13 & 155 & 1033 &  \\
J1619.5$-$5014 & 333.20 & $-$0.02 & 4.74 & 60070.728 & 1367428632 & 4.8 & 66.4 & 6.17 & 156 & 1322 &  \\
J1620.5$-$5729 & 328.20 & $-$5.27 & 5.52 & 60070.729 & 1367428632 & 2.7 & 59.1 & 5.38 & 136 & 371 & Y \\
J1620.8$-$4958 & 333.55 & +0.03 & 4.37 & 60070.729 & 1367428632 & 5.1 & 66.7 & 6.05 & 153 & 1371 &  \\
J1622.2$-$7202 & 317.68 & $-$15.51 & 2.83 & 60069.733 & 1367342464 & 0.4 & 44.6 & 3.00 & 76 & 138 &  \\
J1622.8$-$4454 & 337.37 & +3.36 & 4.43 & 60083.700 & 1368550080 & 4.9 & 71.6 & 6.23 & 158 & 517 &  \\
J1623.7$-$2315 & 353.54 & +18.14 & 4.64 & 60113.613 & 1371131072 & 3.6 & 86.4 & 3.52 & 89 & 75 &  \\
J1624.1$-$4417 & 337.98 & +3.63 & 5.15 & 60083.700 & 1368550080 & 4.2 & 72.3 & 6.21 & 157 & 501 &  \\
J1624.1$-$4941 & 334.12 & $-$0.15 & 5.85 & 60070.731 & 1367428632 & 5.3 & 66.9 & 5.87 & 148 & 1384 &  \\
J1627.7$-$5749 & 328.63 & $-$6.20 & 4.55 & 60070.734 & 1367428632 & 3.0 & 58.8 & 5.51 & 139 & 335 &  \\
J1628.0$-$4920 & 334.82 & $-$0.36 & 5.72 & 60070.734 & 1367428632 & 5.7 & 67.3 & 5.77 & 146 & 1354 &  \\
J1630.1$-$1049 & 4.93 & +24.83 & 3.38 & 60076.719 & 1367946928 & 2.5 & 74.1 & 3.01 & 76 & 123 &  \\
J1630.1$-$3546 & 344.96 & +8.68 & 5.57 & 60083.700 & 1368550080 & 4.7 & 80.8 & 6.27 & 159 & 216 &  \\
J1634.0$-$4742 & 336.69 & +0.03 & 4.70 & 60070.738 & 1367428632 & 7.3 & 68.9 & 6.14 & 155 & 1604 &  \\
J1635.4$-$3249 & 347.93 & +9.85 & 4.54 & 60113.614 & 1371131072 & 6.6 & 83.4 & 3.64 & 92 & 181 &  \\
J1640.3$-$4917 & 336.21 & $-$1.80 & 2.96 & 60070.742 & 1367428632 & 5.7 & 67.3 & 5.64 & 143 & 727 &  \\
J1641.3$-$2908 & 351.61 & +11.30 & 5.53 & 60113.614 & 1371131072 & 4.3 & 85.7 & 3.64 & 92 & 154 &  \\
J1643.3$-$3148 & 349.82 & +9.24 & 3.60 & 60113.614 & 1371131072 & 6.5 & 83.5 & 3.65 & 92 & 189 &  \\
J1644.9$-$4921 & 336.64 & $-$2.41 & 4.22 & 60070.745 & 1367428632 & 5.7 & 67.3 & 5.58 & 141 & 612 &  \\
J1646.5$-$4406 & 340.82 & +0.80 & 3.62 & 60083.711 & 1368550080 & 3.9 & 72.5 & 6.82 & 173 & 1118 &  \\
J1646.7$-$2154 & 358.16 & +14.91 & 4.75 & 60113.614 & 1371131072 & 6.8 & 83.2 & 3.65 & 92 & 117 &  \\
J1647.5$-$5319 & 333.88 & $-$5.30 & 11.92 & 60070.747 & 1367428632 & 1.8 & 63.3 & 5.19 & 131 & 381 & Y \\
J1652.2$-$4516 & 340.57 & $-$0.74 & 5.83 & 60083.715 & 1368550080 & 5.0 & 71.3 & 7.09 & 179 & 1147 &  \\
J1655.9$-$3940 & 345.35 & +2.28 & 4.51 & 60083.717 & 1368550080 & 1.3 & 76.9 & 6.96 & 176 & 770 &  \\
J1656.4$-$0410 & 14.98 & +23.18 & 3.13 & 60076.737 & 1367946928 & 8.9 & 67.5 & 3.68 & 93 & 81 &  \\
J1702.3$-$2525 & 357.52 & +9.94 & 3.29 & 60084.743 & 1368640168 & 7.8 & 82.2 & 7.69 & 194 & 423 &  \\
J1704.0$-$4226 & 344.12 & $-$0.64 & 3.62 & 60083.723 & 1368550080 & 2.3 & 74.2 & 7.29 & 184 & 1238 &  \\
J1704.8$-$4030 & 345.74 & +0.43 & 2.74 & 60083.723 & 1368550080 & 1.0 & 76.1 & 7.37 & 186 & 1397 &  \\
J1704.9$-$4303 & 343.73 & $-$1.13 & 4.19 & 60083.724 & 1368550080 & 2.9 & 73.6 & 7.29 & 184 & 1001 &  \\
J1705.2$-$3850 & 347.13 & +1.36 & 5.69 & 60083.724 & 1368550080 & 1.9 & 77.8 & 7.14 & 181 & 957 &  \\
J1705.4$-$4850 & 339.17 & $-$4.70 & 9.44 & 60070.760 & 1367428632 & 6.2 & 67.8 & 5.21 & 132 & 447 & Y \\
J1708.6$-$4312 & 344.01 & $-$1.76 & 4.51 & 60083.726 & 1368550080 & 3.0 & 73.4 & 7.31 & 185 & 828 &  \\
J1709.4$-$0328 & 17.44 & +20.77 & 4.00 & 60093.718 & 1369415648 & 8.6 & 65.8 & 3.93 & 100 & 93 &  \\
J1710.3$-$3039 & 354.32 & +5.42 & 3.71 & 60084.743 & 1368640168 & 7.1 & 82.9 & 7.35 & 186 & 301 &  \\
J1711.0$-$3002 & 354.91 & +5.64 & 4.21 & 60084.743 & 1368640168 & 6.6 & 83.4 & 7.25 & 183 & 288 & Y$^\dagger$ \\
J1714.8$-$1421 & 8.50 & +13.86 & 5.00 & 60076.749 & 1367946928 & 1.9 & 77.6 & 4.62 & 117 & 102 &  \\
J1717.5$-$4022 & 347.28 & $-$1.46 & 4.80 & 60083.732 & 1368550080 & 1.0 & 76.2 & 7.16 & 181 & 917 &  \\
J1717.5$-$5804 & 332.60 & $-$11.50 & 2.75 & 60070.762 & 1367428632 & 3.4 & 58.6 & 5.22 & 132 & 188 &  \\
J1717.6$-$4404 & 344.26 & $-$3.59 & 5.35 & 60083.732 & 1368550080 & 3.8 & 72.6 & 6.92 & 175 & 572 &  \\
J1717.8$-$3906 & 348.35 & $-$0.77 & 4.32 & 60083.733 & 1368550080 & 1.7 & 77.5 & 7.53 & 191 & 1151 &  \\
J1719.4$-$4242 & 345.57 & $-$3.09 & 5.98 & 60083.734 & 1368550080 & 2.5 & 73.9 & 6.91 & 175 & 629 &  \\
J1721.7$-$3917 & 348.62 & $-$1.49 & 3.72 & 60083.735 & 1368550080 & 1.5 & 77.3 & 7.15 & 181 & 874 &  \\
J1722.1$-$3205 & 354.60 & +2.54 & 4.30 & 60084.743 & 1368640168 & 6.4 & 83.6 & 7.32 & 185 & 603 &  \\
J1725.1$-$1924 & 5.58 & +9.03 & 3.53 & 60076.757 & 1367946928 & 6.6 & 82.6 & 5.18 & 131 & 117 &  \\
J1726.2$-$3207 & 355.06 & +1.79 & 4.24 & 60084.743 & 1368640168 & 6.0 & 84.0 & 7.39 & 187 & 711 &  \\
J1727.4+0326 & 26.23 & +20.23 & 3.21 & 60093.718 & 1369415648 & 3.2 & 59.7 & 3.91 & 99 & 91 &  \\
J1727.7$-$3621 & 351.73 & $-$0.83 & 4.61 & 60083.739 & 1368550080 & 4.2 & 80.3 & 7.21 & 182 & 1019 &  \\
J1729.1$-$3503 & 352.96 & $-$0.33 & 3.61 & 60083.740 & 1368550080 & 5.5 & 81.5 & 7.40 & 187 & 1243 &  \\
J1729.9$-$2952 & 357.39 & +2.37 & 4.11 & 60084.743 & 1368640168 & 3.7 & 86.3 & 7.36 & 186 & 623 &  \\
J1730.1$-$3422 & 353.65 & $-$0.14 & 5.25 & 60083.741 & 1368550080 & 6.1 & 82.2 & 7.46 & 189 & 1310 &  \\
J1730.1$-$4343 & 345.82 & $-$5.27 & 13.61 & 60083.741 & 1368550080 & 3.5 & 72.9 & 7.03 & 178 & 379 & Y \\
J1730.3$-$2913 & 357.96 & +2.67 & 6.06 & 60084.743 & 1368640168 & 3.1 & 86.9 & 7.35 & 186 & 586 & Y \\
J1730.5$-$3543 & 352.57 & $-$0.95 & 5.03 & 60083.741 & 1368550080 & 4.8 & 80.9 & 7.06 & 179 & 953 &  \\
J1730.8$-$3806 & 350.60 & $-$2.31 & 5.08 & 60083.742 & 1368550080 & 2.5 & 78.5 & 6.78 & 172 & 664 &  \\
J1732.0$-$2659 & 0.04 & +3.58 & 5.68 & 60084.743 & 1368640168 & 1.4 & 88.6 & 7.30 & 185 & 443 &  \\
J1733.2$-$2915 & 358.29 & +2.12 & 4.77 & 60084.743 & 1368640168 & 2.9 & 87.1 & 7.47 & 189 & 664 &  \\
J1734.5$-$2818 & 359.25 & +2.39 & 5.09 & 60084.743 & 1368640168 & 2.1 & 87.9 & 7.38 & 187 & 624 &  \\
J1735.1$-$3708 & 351.89 & $-$2.49 & 5.88 & 60083.742 & 1368550080 & 3.5 & 79.4 & 6.72 & 170 & 624 &  \\
J1735.2$-$2153 & 4.76 & +5.71 & 4.71 & 60084.743 & 1368640168 & 5.0 & 85.0 & 7.40 & 187 & 217 & Y \\
J1735.4$-$2944 & 358.14 & +1.44 & 3.83 & 60084.743 & 1368640168 & 3.3 & 86.7 & 7.74 & 196 & 806 &  \\
J1737.1$-$2901 & 358.94 & +1.53 & 8.32 & 60084.743 & 1368640168 & 2.7 & 87.3 & 7.77 & 197 & 967 & Y \\
J1737.3$-$3332 & 355.17 & $-$0.93 & 4.85 & 60083.742 & 1368550080 & 7.1 & 82.9 & 7.05 & 178 & 949 &  \\
J1738.2$-$2510 & 2.34 & +3.37 & 3.67 & 60084.744 & 1368640168 & 2.0 & 88.0 & 7.22 & 183 & 446 &  \\
J1738.8$-$3241 & 356.04 & $-$0.74 & 5.89 & 60084.744 & 1368640168 & 6.1 & 83.9 & 7.75 & 196 & 1039 &  \\
J1739.1$-$1059 & 14.63 & +10.61 & 6.25 & 60076.761 & 1367946928 & 2.9 & 74.1 & 5.51 & 139 & 156 & Y \\
J1739.2$-$2717 & 0.66 & +2.05 & 5.70 & 60084.744 & 1368640168 & 1.3 & 88.7 & 7.41 & 187 & 672 &  \\
J1740.5$-$2554 & 1.99 & +2.55 & 4.49 & 60084.745 & 1368640168 & 1.4 & 88.6 & 7.21 & 182 & 596 &  \\
J1741.1$-$1617 & 10.28 & +7.47 & 4.43 & 60076.761 & 1367946928 & 4.2 & 79.3 & 5.38 & 136 & 216 &  \\
J1741.3$-$3357 & 355.25 & $-$1.86 & 5.74 & 60083.742 & 1368550080 & 6.8 & 82.3 & 6.78 & 171 & 755 &  \\
J1742.8$-$2246 & 4.94 & +3.74 & 5.86 & 60084.747 & 1368640168 & 4.1 & 85.9 & 7.41 & 187 & 368 & Y \\
J1743.1$-$3049 & 358.10 & $-$0.54 & 4.30 & 60084.747 & 1368640168 & 4.3 & 85.7 & 8.36 & 212 & 1155 &  \\
J1743.4$-$2406 & 3.87 & +2.93 & 5.31 & 60084.747 & 1368640168 & 2.9 & 87.1 & 7.20 & 182 & 499 &  \\
J1743.4$-$3123 & 357.66 & $-$0.87 & 3.53 & 60084.747 & 1368640168 & 4.9 & 85.1 & 7.97 & 202 & 988 &  \\
J1744.7$-$1557 & 11.01 & +6.91 & 3.75 & 60076.761 & 1367946928 & 4.5 & 78.8 & 5.37 & 136 & 234 &  \\
J1744.9$-$3322 & 356.14 & $-$2.19 & 5.80 & 60084.748 & 1368640168 & 6.8 & 83.2 & 7.28 & 184 & 679 &  \\
J1745.6$-$3145 & 357.60 & $-$1.48 & 4.62 & 60084.749 & 1368640168 & 5.2 & 84.8 & 7.60 & 192 & 796 &  \\
J1745.6$-$3626 & 353.59 & $-$3.91 & 4.83 & 60083.742 & 1368550080 & 5.0 & 79.7 & 6.98 & 177 & 378 & Y$^\dagger$ \\
J1746.1$-$2541 & 2.84 & +1.58 & 4.18 & 60084.749 & 1368640168 & 1.6 & 88.4 & 7.58 & 192 & 758 &  \\
J1746.4$-$3541 & 354.32 & $-$3.66 & 5.50 & 60083.742 & 1368550080 & 5.7 & 80.4 & 6.86 & 174 & 399 &  \\
J1747.0$-$3505 & 354.89 & $-$3.46 & 3.96 & 60083.742 & 1368550080 & 6.2 & 80.9 & 6.84 & 173 & 427 & Y$^\dagger$ \\
J1747.6+0324 & 28.63 & +15.74 & 2.77 & 60093.725 & 1369415648 & 2.3 & 59.9 & 4.37 & 111 & 120 &  \\
J1747.8$-$3006 & 359.25 & $-$1.03 & 3.57 & 60084.750 & 1368640168 & 3.6 & 86.4 & 8.70 & 220 & 952 &  \\
J1747.9$-$3224 & 357.29 & $-$2.23 & 4.37 & 60084.750 & 1368640168 & 5.9 & 84.1 & 7.30 & 185 & 644 &  \\
J1749.6$-$3143 & 358.06 & $-$2.18 & 5.84 & 60084.752 & 1368640168 & 5.2 & 84.8 & 7.32 & 185 & 662 &  \\
J1750.0$-$3347 & 356.33 & $-$3.33 & 4.81 & 60084.752 & 1368640168 & 7.2 & 82.8 & 7.42 & 188 & 458 &  \\
J1750.4$-$3023 & 359.29 & $-$1.66 & 5.72 & 60084.752 & 1368640168 & 3.9 & 86.1 & 7.75 & 196 & 755 &  \\
J1750.6$-$1906 & 9.02 & +4.08 & 4.39 & 60076.761 & 1367946928 & 7.8 & 81.1 & 5.48 & 139 & 395 &  \\
J1750.8$-$1246 & 14.54 & +7.22 & 9.69 & 60076.761 & 1367946928 & 4.7 & 75.3 & 5.34 & 135 & 225 & Y \\
J1752.3$-$2914 & 0.48 & $-$1.41 & 4.30 & 60084.754 & 1368640168 & 2.8 & 87.2 & 8.00 & 202 & 770 &  \\
J1752.4$-$0758 & 18.94 & +9.28 & 13.70 & 60076.761 & 1367946928 & 7.2 & 70.6 & 5.53 & 140 & 177 & Y \\
J1753.2$-$2848 & 0.96 & $-$1.36 & 5.48 & 60084.754 & 1368640168 & 2.4 & 87.6 & 7.93 & 201 & 771 &  \\
J1754.8$-$3200 & 358.38 & $-$3.27 & 5.14 & 60084.755 & 1368640168 & 5.4 & 84.6 & 7.34 & 186 & 484 &  \\
J1755.9$-$4009 & 351.39 & $-$7.52 & 6.88 & 60083.742 & 1368550080 & 4.7 & 75.6 & 6.83 & 173 & 194 & Y \\
J1757.1$-$2848 & 1.41 & $-$2.11 & 5.65 & 60084.757 & 1368640168 & 2.4 & 87.6 & 7.38 & 187 & 597 &  \\
J1757.1$-$3554 & 355.23 & $-$5.63 & 5.75 & 60083.742 & 1368550080 & 6.8 & 79.2 & 6.80 & 172 & 223 &  \\
J1757.4$-$3125 & 359.16 & $-$3.46 & 4.90 & 60084.757 & 1368640168 & 4.9 & 85.1 & 7.40 & 187 & 434 &  \\
J1758.0$-$2953 & 0.56 & $-$2.81 & 4.88 & 60084.758 & 1368640168 & 3.4 & 86.6 & 7.28 & 184 & 520 &  \\
J1758.6$-$2404 & 5.66 & $-$0.04 & 2.67 & 60084.758 & 1368640168 & 2.9 & 87.1 & 8.30 & 210 & 1225 &  \\
J1758.8$-$2326 & 6.24 & +0.25 & 4.07 & 60084.758 & 1368640168 & 3.5 & 86.5 & 8.32 & 211 & 1273 &  \\
J1759.7$-$2141 & 7.85 & +0.94 & 2.49 & 60084.759 & 1368640168 & 5.2 & 84.8 & 7.76 & 196 & 913 &  \\
J1759.8$-$1250 & 15.58 & +5.30 & 3.80 & 60076.761 & 1367946928 & 6.9 & 74.6 & 5.48 & 139 & 314 &  \\
J1800.3$-$5237 & 340.54 & $-$14.07 & 4.31 & 60104.739 & 1370367808 & 7.6 & 62.5 & 2.81 & 71 & 138 &  \\
J1800.5$-$2910 & 1.45 & $-$2.92 & 5.42 & 60084.759 & 1368640168 & 2.8 & 87.2 & 7.24 & 183 & 486 &  \\
J1801.0$-$2802 & 2.49 & $-$2.47 & 5.17 & 60084.760 & 1368640168 & 1.8 & 88.2 & 7.24 & 183 & 527 &  \\
J1801.1$-$2626 & 3.90 & $-$1.70 & 3.46 & 60084.760 & 1368640168 & 1.2 & 88.8 & 7.39 & 187 & 653 &  \\
J1802.1$-$2652 & 3.63 & $-$2.11 & 5.83 & 60084.760 & 1368640168 & 1.1 & 88.9 & 7.27 & 184 & 575 &  \\
J1802.4$-$3041 & 0.33 & $-$4.04 & 2.78 & 60084.761 & 1368640168 & 4.2 & 85.8 & 7.51 & 190 & 330 & Y$^\dagger$ \\
J1805.1$-$3618 & 355.66 & $-$7.24 & 3.11 & 60083.742 & 1368550080 & 7.8 & 78.0 & 6.73 & 170 & 163 &  \\
J1805.9$-$1549 & 13.70 & +2.55 & 12.58 & 60076.761 & 1367946928 & 8.8 & 76.4 & 5.37 & 136 & 575 & Y \\
J1808.4$-$3522 & 356.82 & $-$7.40 & 3.33 & 60099.747 & 1369936552 & 8.4 & 78.7 & 3.94 & 100 & 156 &  \\
J1809.2$-$2726 & 3.90 & $-$3.76 & 5.44 & 60084.765 & 1368640168 & 1.4 & 88.6 & 7.23 & 183 & 350 &  \\
J1809.5$-$0816 & 20.75 & +5.41 & 5.70 & 60093.741 & 1369415648 & 10.0 & 71.5 & 5.20 & 131 & 351 &  \\
J1810.3+1332 & 40.64 & +15.13 & 5.40 & 60097.753 & 1369764224 & 9.3 & 49.0 & 3.64 & 92 & 114 &  \\
J1810.8$-$3347 & 358.45 & $-$7.08 & 5.74 & 60084.766 & 1368640168 & 7.2 & 82.8 & 7.20 & 182 & 160 &  \\
J1813.7$-$1152 & 18.08 & +2.79 & 4.43 & 60094.761 & 1369505736 & 7.1 & 73.6 & 6.07 & 153 & 543 &  \\
J1814.2$-$1012 & 19.59 & +3.48 & 9.46 & 60094.761 & 1369505736 & 7.5 & 72.2 & 6.04 & 153 & 469 & Y \\
J1814.2$-$3000 & 2.16 & $-$5.95 & 5.93 & 60084.769 & 1368640168 & 3.5 & 86.5 & 6.91 & 175 & 191 &  \\
J1814.7$-$3420 & 358.35 & $-$8.06 & 3.56 & 60084.769 & 1368640168 & 7.7 & 82.3 & 7.31 & 185 & 139 &  \\
J1815.8$-$1416 & 16.21 & +1.19 & 5.09 & 60094.761 & 1369505736 & 6.6 & 76.0 & 6.15 & 156 & 901 & Y$^\dagger$ \\
J1816.4$-$2727 & 4.65 & $-$5.17 & 4.70 & 60084.770 & 1368640168 & 1.4 & 88.6 & 7.02 & 178 & 229 &  \\
J1816.7+1749 & 45.34 & +15.51 & 3.16 & 60097.753 & 1369764224 & 6.2 & 45.0 & 3.59 & 91 & 107 &  \\
J1817.9$-$1135 & 18.80 & +2.02 & 5.86 & 60094.761 & 1369505736 & 6.2 & 73.8 & 6.22 & 157 & 670 &  \\
J1817.9$-$3334 & 359.34 & $-$8.28 & 3.85 & 60084.771 & 1368640168 & 7.0 & 83.0 & 7.52 & 190 & 133 & Y$^\dagger$ \\
J1818.1$-$2000 & 11.41 & $-$2.00 & 3.49 & 60084.771 & 1368640168 & 6.8 & 83.2 & 6.89 & 174 & 659 &  \\
J1818.8$-$0828 & 21.67 & +3.30 & 5.27 & 60094.761 & 1369505736 & 7.4 & 70.9 & 6.01 & 152 & 480 &  \\
J1819.9$-$2926 & 3.25 & $-$6.78 & 4.99 & 60084.773 & 1368640168 & 3.0 & 87.0 & 6.62 & 168 & 165 & Y \\
J1820.3$-$1009 & 20.35 & +2.18 & 5.76 & 60094.761 & 1369505736 & 6.2 & 72.6 & 5.95 & 150 & 637 &  \\
J1820.6$-$0326 & 26.34 & +5.26 & 4.41 & 60093.748 & 1369415648 & 5.2 & 66.7 & 5.43 & 137 & 347 &  \\
J1824.7+2403 & 52.06 & +16.28 & 3.92 & 60097.753 & 1369764224 & 7.0 & 39.0 & 3.62 & 92 & 98 &  \\
J1825.2+0715 & 36.47 & +9.11 & 10.01 & 60093.752 & 1369415648 & 5.8 & 56.0 & 5.53 & 140 & 199 & Y \\
J1825.9$-$3153 & 1.62 & $-$9.03 & 5.57 & 60084.777 & 1368640168 & 5.3 & 84.7 & 6.75 & 171 & 122 &  \\
J1826.2$-$2830 & 4.71 & $-$7.59 & 18.28 & 60084.777 & 1368640168 & 2.2 & 87.8 & 6.37 & 161 & 147 & Y \\
J1828.2$-$3252 & 0.95 & $-$9.90 & 5.00 & 60084.778 & 1368640168 & 6.3 & 83.7 & 6.71 & 170 & 112 &  \\
J1830.7$-$1634 & 15.85 & $-$3.04 & 2.23 & 60094.761 & 1369505736 & 4.7 & 79.4 & 5.88 & 149 & 518 &  \\
J1830.7$-$2414 & 9.01 & $-$6.55 & 5.65 & 60084.780 & 1368640168 & 2.8 & 87.2 & 6.37 & 161 & 190 &  \\
J1831.2$-$0013 & 30.43 & +4.38 & 5.15 & 60093.756 & 1369415648 & 2.3 & 63.5 & 5.69 & 144 & 364 &  \\
J1831.4$-$2909 & 4.64 & $-$8.89 & 14.80 & 60084.781 & 1368640168 & 2.7 & 87.3 & 6.25 & 158 & 125 & Y \\
J1832.4$-$0847 & 22.94 & +0.19 & 3.77 & 60094.761 & 1369505736 & 5.1 & 71.9 & 6.47 & 164 & 1541 &  \\
J1833.4+1109 & 40.92 & +9.01 & 5.21 & 60097.753 & 1369764224 & 7.9 & 52.0 & 3.58 & 91 & 184 &  \\
J1834.9$-$2819 & 5.75 & $-$9.21 & 5.42 & 60084.783 & 1368640168 & 2.0 & 88.0 & 6.17 & 156 & 123 &  \\
J1836.1+1143 & 41.72 & +8.68 & 3.91 & 60097.753 & 1369764224 & 7.2 & 51.5 & 3.56 & 90 & 188 &  \\
J1836.1$-$2656 & 7.11 & $-$8.85 & 17.27 & 60084.784 & 1368640168 & 1.2 & 88.8 & 6.00 & 152 & 131 & Y \\
J1836.8$-$0727 & 24.65 & $-$0.18 & 4.57 & 60094.761 & 1369505736 & 5.9 & 70.6 & 6.58 & 166 & 1525 &  \\
J1838.4$-$0630 & 25.65 & $-$0.09 & 3.46 & 60094.761 & 1369505736 & 6.7 & 69.7 & 6.49 & 164 & 1544 &  \\
J1840.4$-$1139 & 21.31 & $-$2.88 & 4.97 & 60094.761 & 1369505736 & 2.0 & 74.9 & 5.88 & 149 & 526 &  \\
J1843.3$-$1242 & 20.69 & $-$4.00 & 4.88 & 60094.762 & 1369505736 & 1.3 & 75.9 & 6.09 & 154 & 427 &  \\
J1843.6$-$2219 & 12.06 & $-$8.35 & 5.66 & 60084.785 & 1368640168 & 4.8 & 85.2 & 5.85 & 148 & 161 &  \\
J1844.8$-$0957 & 23.32 & $-$3.07 & 5.78 & 60094.762 & 1369505736 & 3.4 & 73.2 & 5.81 & 147 & 498 &  \\
J1847.2$-$0141 & 30.95 & +0.16 & 5.62 & 60093.760 & 1369415648 & 4.3 & 64.8 & 6.20 & 157 & 1387 &  \\
J1847.2$-$0200 & 30.67 & +0.00 & 3.89 & 60093.760 & 1369415648 & 4.5 & 65.1 & 6.26 & 158 & 1417 &  \\
J1847.7$-$3433 & 1.11 & $-$14.30 & 5.37 & 60099.751 & 1369936552 & 6.0 & 82.1 & 3.74 & 95 & 81 &  \\
J1849.1$-$0652 & 26.55 & $-$2.62 & 5.36 & 60094.765 & 1369505736 & 6.3 & 70.1 & 5.80 & 147 & 544 &  \\
J1851.0+1558 & 47.18 & +7.29 & 5.39 & 60097.758 & 1369764224 & 3.0 & 47.3 & 3.55 & 90 & 200 &  \\
J1851.5+0718 & 39.46 & +3.30 & 4.96 & 60093.760 & 1369415648 & 6.8 & 55.8 & 5.62 & 142 & 371 &  \\
J1852.7$-$1242 & 21.72 & $-$6.04 & 5.80 & 60094.768 & 1369505736 & 1.3 & 75.9 & 5.67 & 143 & 308 &  \\
J1853.0$-$0956 & 24.24 & $-$4.88 & 5.40 & 60094.768 & 1369505736 & 3.4 & 73.2 & 5.60 & 142 & 363 &  \\
J1853.6$-$0620 & 27.53 & $-$3.39 & 7.53 & 60094.769 & 1369505736 & 6.8 & 69.6 & 5.55 & 140 & 445 & Y \\
J1854.1+0142 & 34.76 & +0.19 & 4.41 & 60093.760 & 1369415648 & 4.2 & 61.2 & 6.06 & 153 & 1191 &  \\
J1854.3$-$3640 & 359.62 & $-$16.35 & 4.45 & 60099.756 & 1369936552 & 3.9 & 80.0 & 3.51 & 89 & 75 &  \\
J1855.7+0224 & 35.57 & +0.14 & 5.50 & 60093.760 & 1369415648 & 4.7 & 60.5 & 5.99 & 151 & 1160 &  \\
J1855.8+0150 & 35.07 & $-$0.14 & 4.82 & 60093.760 & 1369415648 & 4.6 & 61.0 & 6.13 & 155 & 1190 &  \\
J1857.8$-$3220 & 4.07 & $-$15.36 & 4.58 & 60084.785 & 1368640168 & 7.3 & 82.7 & 6.00 & 152 & 82 &  \\
J1858.5$-$4446 & 351.94 & $-$19.92 & 4.61 & 60099.759 & 1369936552 & 4.4 & 71.9 & 3.32 & 84 & 89 &  \\
J1859.2$-$0706 & 27.47 & $-$4.96 & 4.61 & 60094.772 & 1369505736 & 6.1 & 70.3 & 5.42 & 137 & 345 &  \\
J1859.8+0411 & 37.62 & +0.06 & 5.31 & 60093.760 & 1369415648 & 6.2 & 58.6 & 5.99 & 151 & 1099 &  \\
J1900.4$-$4941 & 347.04 & $-$21.74 & 4.97 & 60104.746 & 1370367808 & 5.4 & 67.0 & 2.62 & 66 & 98 &  \\
J1901.5$-$2208 & 14.04 & $-$12.01 & 4.73 & 60084.785 & 1368640168 & 7.3 & 82.7 & 5.93 & 150 & 121 &  \\
J1901.8$-$0718 & 27.59 & $-$5.62 & 3.86 & 60094.774 & 1369505736 & 5.9 & 70.5 & 5.27 & 133 & 316 &  \\
J1902.5+0654 & 40.35 & +0.68 & 2.85 & 60093.760 & 1369415648 & 8.3 & 55.8 & 5.77 & 146 & 855 &  \\
J1904.0$-$5925 & 336.91 & $-$24.68 & 5.65 & 60104.749 & 1370367808 & 4.5 & 57.3 & 2.57 & 65 & 83 &  \\
J1905.2$-$5120 & 345.54 & $-$22.93 & 3.88 & 60104.750 & 1370367808 & 3.8 & 65.4 & 2.55 & 64 & 91 &  \\
J1906.0$-$1718 & 18.96 & $-$10.95 & 4.63 & 60094.777 & 1369505736 & 4.5 & 80.5 & 4.90 & 124 & 153 &  \\
J1906.4$-$1757 & 18.41 & $-$11.30 & 3.37 & 60094.778 & 1369505736 & 5.1 & 81.1 & 4.81 & 122 & 148 &  \\
J1906.4$-$2001 & 16.49 & $-$12.17 & 5.96 & 60094.778 & 1369505736 & 7.1 & 83.2 & 4.86 & 123 & 138 &  \\
J1908.7+0812 & 42.20 & $-$0.08 & 4.10 & 60097.771 & 1369764224 & 10.5 & 55.0 & 3.72 & 94 & 883 &  \\
J1911.1$-$3221 & 5.15 & $-$17.98 & 5.24 & 60099.767 & 1369936552 & 8.2 & 84.2 & 2.96 & 75 & 77 &  \\
J1911.4$-$4856 & 348.34 & $-$23.26 & 5.96 & 60104.754 & 1370367808 & 6.2 & 67.8 & 2.45 & 62 & 90 &  \\
J1912.7+0957 & 44.21 & $-$0.15 & 5.40 & 60097.773 & 1369764224 & 8.8 & 53.3 & 3.70 & 94 & 859 &  \\
J1914.6$-$1157 & 24.79 & $-$10.52 & 4.64 & 60094.783 & 1369505736 & 1.7 & 75.1 & 4.43 & 112 & 164 &  \\
J1916.8$-$3025 & 7.52 & $-$18.42 & 1.87 & 60105.776 & 1370457464 & 7.8 & 82.2 & 2.39 & 60 & 78 &  \\
J1917.6$-$3204 & 5.96 & $-$19.16 & 5.05 & 60099.772 & 1369936552 & 8.5 & 84.5 & 2.76 & 70 & 76 &  \\
J1920.0$-$2622 & 11.75 & $-$17.58 & 4.16 & 60105.776 & 1370457464 & 6.3 & 83.7 & 2.37 & 60 & 86 &  \\
J1920.8+0418 & 40.14 & $-$4.56 & 4.97 & 60139.684 & 1373387040 & 7.4 & 58.2 & 3.14 & 79 & 296 &  \\
J1940.2$-$2511 & 14.69 & $-$21.36 & 4.93 & 60105.776 & 1370457464 & 2.5 & 87.5 & 2.38 & 60 & 91 &  \\
J1946.5$-$1527 & 24.86 & $-$19.01 & 5.12 & 60094.803 & 1369505736 & 2.9 & 78.6 & 2.93 & 74 & 101 &  \\
J1947.0+0031 & 39.84 & $-$12.10 & 3.91 & 60139.684 & 1373387040 & 1.7 & 62.7 & 3.14 & 79 & 143 &  \\
J1949.2$-$1453 & 25.70 & $-$19.39 & 3.96 & 60094.803 & 1369505736 & 2.7 & 78.0 & 2.93 & 74 & 99 &  \\
J1957.6+1230 & 51.78 & $-$8.55 & 4.01 & 60097.794 & 1369764224 & 7.1 & 50.5 & 3.12 & 79 & 170 &  \\
J2007.2$-$3157 & 9.64 & $-$29.11 & 3.71 & 60105.790 & 1370457464 & 5.3 & 84.7 & 2.08 & 52 & 66 &  \\
J2046.1$-$1626 & 30.08 & $-$32.61 & 4.40 & 60118.786 & 1371581520 & 4.0 & 79.4 & 1.62 & 41 & 57 &  \\
J2056.4$-$5922 & 336.95 & $-$38.88 & 4.63 & 58380.571 & 1221399680 & 6.2 & 56.9 & 1.61 & 41 & 49 &  \\
J2241.4$-$8327 & 307.10 & $-$32.58 & 3.31 & 58445.505 & 1227009976 & 12.1 & 32.8 & 1.33 & 34 & 56 &  \\
J2342.4$-$4739 & 331.64 & $-$65.43 & 3.89 & 58430.525 & 1225713560 & 7.2 & 69.1 & 1.24 & 31 & 30 &  \\

\end{longtable}

\twocolumn

\end{document}